\documentclass[prc,letterpaper,twocolumn,showpacs,showkeys,lengthcheck,tightenlines,
               nofootinbib,preprintnumbers,superscriptaddress]{revtex4-1} 

\pdfpageattr {/Group << /S /Transparency /I true /CS /DeviceRGB>>}

\usepackage[utf8]{inputenc}
\usepackage{newtxtext,newtxmath}

\usepackage{dcolumn}
\usepackage{bm}
\usepackage{amsmath,amssymb,amsfonts}

\usepackage{graphicx}

\usepackage{mathrsfs}   % \mathscr
\usepackage{graphicx}   % figures
\usepackage{bbold}      % bold font

\usepackage{color}      % color is used in text
\usepackage{footnote}   % footnote in tabular environment, must be loaded after color

\usepackage{slashed}    % Feynman slash

\usepackage{rotating}   % to rotate tables
\usepackage{multirow}   % multicolumn and multirow

\usepackage[svgnames]{xcolor}
\usepackage[english]{babel}
\usepackage{blindtext}
\usepackage{microtype}
\usepackage{tikz}

\usepackage[colorlinks=true,linkcolor=blue,urlcolor=blue,citecolor=blue]{hyperref}
\usepackage{ifpdf}

\def\bea{\begin{eqnarray}}
\def\eea{\end{eqnarray}}

\newcommand{\gsim}{\gtrsim}

\def\e{\epsilon}
\def\ve{\varepsilon}

\newcommand{\eq}[1]{Eq.~(\ref{#1})}

\def\hs{\hspace}

\def\no{\nonumber}

\def\lf{\left}
\def\rg{\right}

\newcommand{\sh}[1]{\slashed{#1}}

\graphicspath{{.}{figures/}} 
\begin{document}

\preprint{NT@UW-18-02}
\title{Eta Decay and Muonic Puzzles}

\author{Yu-Sheng Liu}
\email{mestelqure@gmail.com}
\affiliation{Tsung-Dao Lee Institute, Shanghai Jiao Tong University, Shanghai 200240, China}
 
\author{Ian C. Clo\"et}
\email{icloet@anl.gov}
\affiliation{Physics Division, Argonne National Laboratory, Argonne, Illinois 60439, U.S.A.}

\author{Gerald A. Miller}
\email{miller@phys.washington.edu}
\affiliation{Department of Physics, University of Washington, Seattle, Washington 98195-1560, U.S.A.}

\preprint{}
%\date{\today}

\begin{abstract}
New physics motivated by muonic puzzles (proton radius and muon $g-2$ discrepancies) is  studied.  Using a light scalar boson $\phi$, assuming Yukawa interactions, accounts for  these muonic puzzles simultaneously. Our previous work limits the existence of such a scalar boson's mass $m_\phi$ from about 160 keV to 60 MeV. We improve this result by including  the influence of all of the possible particles that couple to the $\phi$ in computing the decay rate.
Doing this involves including the strong interaction physics, involving quarks,  necessary to compute the $\eta\pi\phi$ vertex function. The  Nambu-Jona-Lasinio model, which accounts for the spontaneous symmetry breaking that yields the constituent mass  is employed to represent the relevant strong-interaction physics.  We use  the $\eta\pi\phi$ vertex function to  reanalyze the electron beam dump experiments. The result is that  the allowed range of $m_\phi$ lies between about 160 keV and 3.5 MeV. This narrow range represents an inviting target for ruling out  or discovering this scalar boson. A possible UV completion of our phenomenological model is  discussed.
\end{abstract}

\maketitle
%===============================================================================
%===============================================================================
\section{Introduction}
The proton charge radius measured using the Lamb shift in muonic hydrogen, $r_p = 0.84087(39)\,$fm~\cite{Pohl:2010zza,Antognini:1900ns}, differs from the CODATA average obtained from hydrogen spectroscopy and $e-p$ scattering,  $r_p = 0.8751(61)\,$fm~\cite{Mohr:2015ccw}, by more than $5\sigma$. Although the discrepancy may arise from subtle lepton-nucleon non-perturbative effects within the Standard Model (SM), or experimental uncertainties \cite{Pohl:2013yb,Carlson:2015jba}, it could also be a signal of new physics involving a violation of lepton universality~\cite{Carlson:2012pc,Carlson:2013mya}. The muon anomalous magnetic moment provides another potential signal of new physics \cite{Lindner:2016bgg}. The BNL measurement \cite{Blum:2013xva} differs from the SM prediction by more than $3\sigma$, $\Delta a_\mu=a_\mu^{\rm exp}-a_\mu^{\rm th}=287(80)\times 10^{-11}$~\cite{Davier:2010nc,Hagiwara:2011af}.

A new scalar boson $\phi$ (we have concluded in our previous work \cite{Liu:2017bzj} that other spin-0 and spin-1 bosons are ruled out), which couples to the muon and proton could explain both the proton radius and $(g-2)_\mu$ puzzles \cite{TuckerSmith:2010ra,Liu:2015sba}. Phenomenological motivation for such a scalar boson as a Higgs portal in the dark sector has been considered theoretically~\cite{Kinoshita:1990aj,Batell:2011qq,Schmidt-Hoberg:2013hba,Essig:2013lka,Knapen:2015hia,Chen:2015vqy,Batell:2016ove,Batell:2017kty} and experimentally~\cite{Izaguirre:2014cza,Chen:2017awl,Chen:2018vkr,Kahn:2018cqs,Berlin:2018pwi}. 
We investigate the couplings of this boson to SM fermions, $\psi$, which appear as Yukawa terms in the effective 
Lagrangian, $\mathcal{L} \supset e\,\epsilon_f\,\phi\,\bar{\psi}_f\,\psi_f$, where $\epsilon_f=g_f/e$, $e$ is the electric charge of the proton, and $f$ is the flavor index. Other authors have pursued this idea, but made further assumptions relating the couplings to different particle species, mass range, etc. We make no {\it a priori} assumptions regarding signs or magnitudes of the coupling constants. The Lamb shift in muonic hydrogen fixes $\epsilon_\mu$ and $\epsilon_p$ to have the same sign. Without loss of generality, we take both $\epsilon_\mu$ and $\epsilon_p$ to be positive, and $\epsilon_e$ and $\epsilon_n$ are allowed to have either sign.

Coupling a single scalar to
up and down quarks in an effective Lagrangian at the MeV scale is not
consistent with the $SU(2)_L \times  U(1)_Y$ gauge invariance of the Standard %Batell:2017kty,
Model above the electroweak symmetry breaking scale.   An ultraviolet (UV) completion is needed. As is also well known~\cite{Pospelov:2017kep},  while it is difficult to create a viable  model of dark scalars with masses in the MeV  range, 
interesting attempts have been made~\cite{Chen:2015vqy,Batell:2017kty}.  %These are used below to defend our approach.

Electron beam dump experiments have been aimed at searching for new particles \cite{Liu:2016mqv,Liu:2017htz,Essig:2013lka,Bjorken:2009mm,Andreas:2012mt}. The typical setup of an electron beam dump experiment involves a  beam stopped by a large amount of material.   The ensuing interactions could produce  new particles via a bremsstrahlung-like process. Such particles would pass through a shield region and decay. These new particles can be detected by their decay products, electron and/or photon pairs, measured by the detector downstream of the decay region. In our previous work, it was assumed that the new particle only couples to electrons. In our simple model, considering the $\phi$ couplings to other SM particles could  dramatically change the exclusion range. 

It is worthwhile to study the production of a new scalar boson by
eta decay.  This is because there are no selection rules  preventing $\phi$ emission and possible complications involving strangeness are absent in many channels.
We will show that $\eta\to\pi^0\bar{\psi}_f\psi_f$ and $\eta\to\pi^0\gamma\gamma$ decay channels are particularly useful. Eta decay to the $\pi^0\bar{\psi}_f\psi_f$ final state is forbidden at tree level in the SM by charge conjugation symmetry, but it is allowed by a virtual $\phi$ emission. Eta decay to $\pi^0\gamma\gamma$ is observed, and the existence of the $\phi$ may open up new channels, whose decay rate should not exceed the observed value.  We will use the Nambu--Jona-Lasinio (NJL) model~\cite{Nambu:1961fr,Nambu:1961tp,Vogl:1991qt,Klevansky:1992qe}, a chiral effective theory of QCD  exhibiting dynamical chiral symmetry breaking, to  provide the strong-interaction input necessary to predict these decay rates. The NJL model satisfies the soft-pion theorems making it an ideal tool with which to determine the coupling of a scalar boson to the Goldstone bosons. Therefore, using the current $\eta$ decay data, we will significantly  improve the constraints on the new scalar boson.

A recent experiment  extracts the proton radius to be $r_p = 0.8335(95)\,$fm~\cite{Beyer79} by measuring the $2S-4P$ transition frequency in electronic  hydrogen. This result agrees with the previous muonic hydrogen experiments \cite{Pohl:2010zza,Antognini:1900ns} but is more than 3 standard deviations away from the CODATA value \cite{Mohr:2015ccw} that is dominated by many previous hydrogen spectroscopy experiments. Three possible scenarios can immediately be envisioned:
\begin{enumerate}
\item the proton radius puzzle is solved,
\item it is too early to use the new experiment as a replacement for many others,
\item new physics may coexist with the CODATA value and the new experiment result.
\end{enumerate}
It is tempting to accept the first scenario, however it defies the results of decades of the electron-proton scattering experiments. On the other hand, the preliminary nuclei radii from laser spectroscopy of $\mu\,{}^4{\rm He}^+$ and $\mu\,{}^3{\rm He}^+$ \cite{Antognini} agree with the electron-nucleus experiments \cite{Sick:2015spa,Sick:2014yha}. The PRad experiment \cite{Gasparian:2017cgp,Gasparian:2014rna,Meziane:2013yma} may shed some light on this direction. For the second approach, one may  argue that the measurement of $2S-4P$ transition frequency is very difficult because of quantum interference effects that involve the details of the experimental setup. It is desirable to have a second experiment on  regular hydrogen for this transition. Moreover, a more recent electron hydrogen experiment~\cite{newH} on the ${}^1\!S-{}^3\!S$ transition finds a radius in agreement with the CODATA value and earlier hydrogen spectroscopy measurements.
 For the third approach, the true value of proton radius may lie within 3 standard deviations of the new experiments and the old CODATA value, and the muonic hydrogen experiments still signal new physics. In other words, the existence of a new scalar meson may not conflict with any of  the experiments.  We examine the latter two possibilities here.

This paper is organized as follows: Sec.~\ref{sec:couplings}  discusses the Lagrangian, introducing $\phi$ couplings to $u$ and $d$ quarks. A possible UV completion is discussed. The $\eta\pi^0\phi$ vertex is discussed in Sect.~\ref{sec:etapiphi} .  Sec.~\ref{sec:decay rate}  presents the $\phi$ and $\eta$ decay rates. Sec.~\ref{sec:beam dump experiments} revisits the beam dump experiments. Sec.~\ref{sec:eta to pi f f} and Sec.~\ref{sec:eta to pi gamma gamma} show the new exclusion region obtained by different $\eta$ decay channels. Sec.~\ref{sec:coexistence} discusses third scenario which the new physics coexists with the new regular hydrogen experiments and the old CODATA value. A conclusion is given in Sec. \ref{sec:conclusion}.

%===============================================================================
%===============================================================================
\section{Lagrangian}\label{sec:couplings}
In our  previous work, the Lagrangian involved interactions  between the $\phi$ and nucleons. This is not sufficient to study effects involving mesons. Coupling between the  $\phi$ and quarks is examined here. 
%===============================================================================
\begin{figure}[tbp]
\centering\includegraphics[width=\columnwidth]{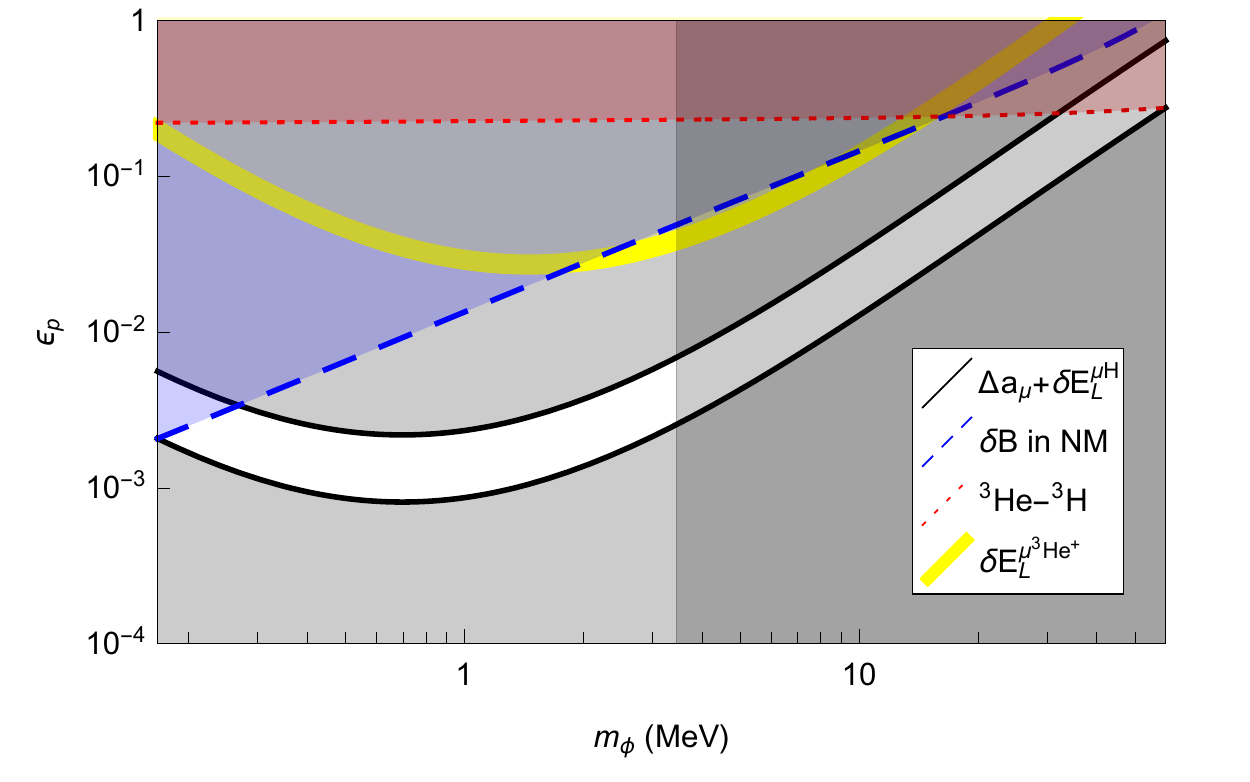}
\caption{\label{fig:gp_exclusion} Exclusion plot for $\epsilon_p$ (shaded region is excluded). The solid black, dotted red, and dashed blue lines are from our previous work \cite{Liu:2016qwd,Liu:2017bzj} corresponding to combining muonic hydrogen \cite{Pohl:2010zza,Antognini:1900ns,Mohr:2015ccw} with muon $g-2$ experiments \cite{Blum:2013xva,Davier:2010nc,Hagiwara:2011af}, the binding energy difference of $^3{\rm He}$ and $^3{\rm H}$ \cite{Friar:1969zz,Friar:1978mr,Coon:1987kt,Miller:1990iz,Wiringa:2013fia,Sick:2001rh,Juster:1985sd,Mccarthy:1977vd}, and the binding energy of nuclear matter per nucleon \cite{Mattuck:1976xt}. The thick yellow solid curve is from the preliminary muonic $^3{\rm He}$ ion laser spectroscopy experiment \cite{Antognini,Sick:2015spa,Sick:2014yha,Franke:2017tpc} combining with the $\epsilon_n$ constraint in Fig. \ref{fig:Rnp_exclusion}. The vertical line indicates the allowed mass range obtained in Fig. \ref{fig:eta_to_pi_gamma_gamma}.}
\end{figure}
%===============================================================================

%===============================================================================
\begin{figure}[tbp]
\centering
\includegraphics[width=\columnwidth]{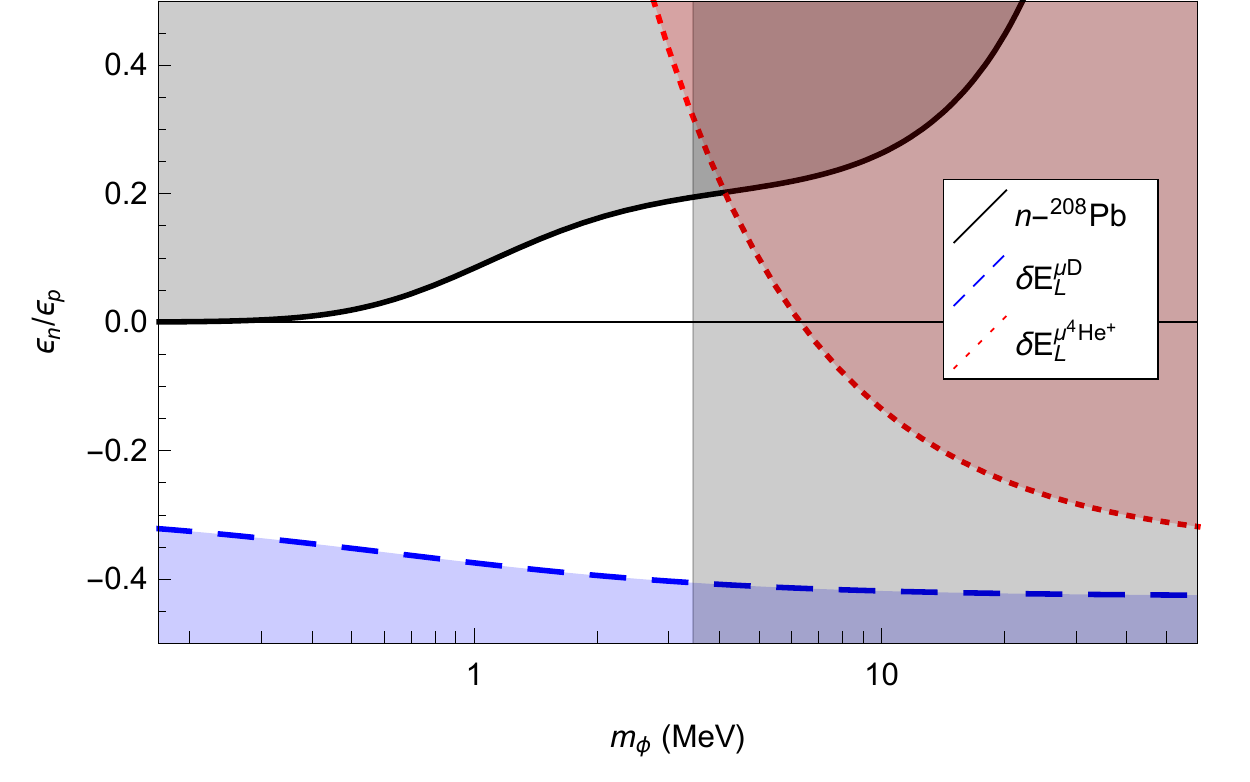}
\caption{\label{fig:Rnp_exclusion} Exclusion plot for $\epsilon_n/\epsilon_p$ (shaded region is excluded). Since $\epsilon_n$ can take either sign, we present $\epsilon_n$ as a ratio to $\epsilon_p$. The solid black, dotted red, and dashed blue lines are from our previous work \cite{Liu:2016qwd,Liu:2017bzj} corresponding to the low energy scattering of neutron on $^{208}{\rm Pb}$ \cite{Leeb:1992qf}, the preliminary muonic $^4{\rm He}$ ion laser spectroscopy experiment \cite{Antognini,Sick:2014yha,Diepold:2016cxv}, and the laser spectroscopy experiment of muonic deuterium \cite{Pohl1:2016xoo,Krauth:2015nja,Antognini:2015moa}. The vertical line indicates the allowed mass range obtained in Fig. \ref{fig:eta_to_pi_gamma_gamma}.
}
\end{figure}
%===============================================================================

%===============================================================================
\begin{figure}[tbp]
\centering
\includegraphics[width=\columnwidth]{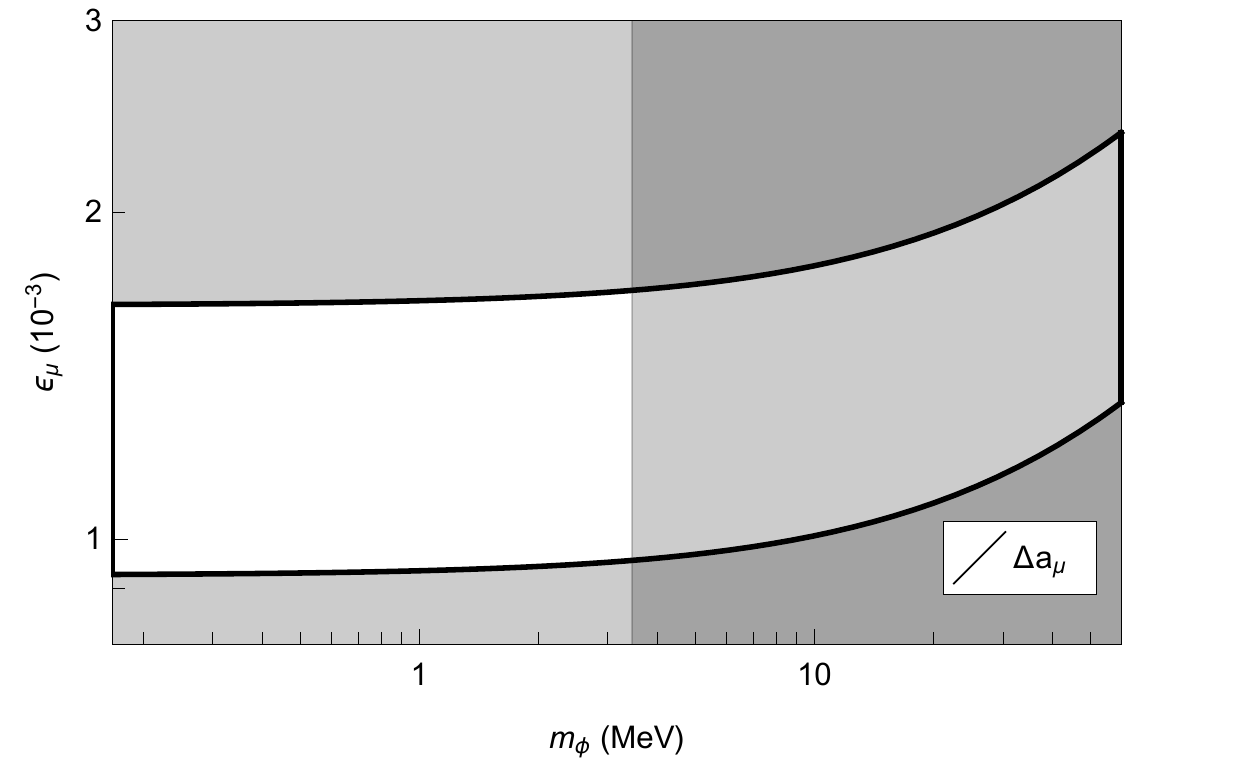}
\caption{\label{fig:gmu_exclusion} Exclusion plot for $\epsilon_\mu$ (shaded region is excluded). The black line uses muon $g-2$ experiments \cite{Blum:2013xva,Davier:2010nc,Hagiwara:2011af}. The vertical line indicates the allowed mass range obtained in Fig. \ref{fig:eta_to_pi_gamma_gamma}.
}
\end{figure}
%===============================================================================

\subsection{$\phi$ couplings to $u$ and $d$ quarks}

\begin{figure}
\centering
\includegraphics[width=\columnwidth]{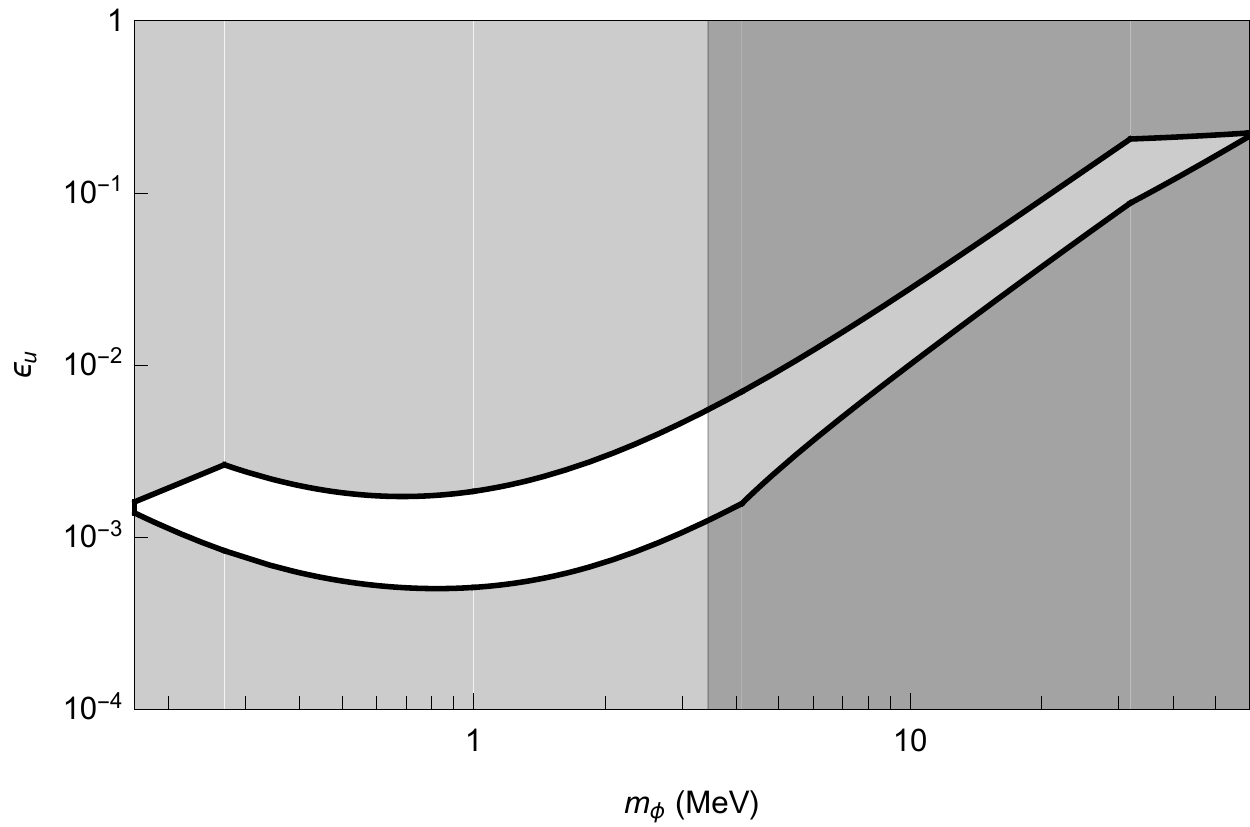}
\caption{\label{fig:gu_exclusion} Exclusion plot for $\epsilon_u$ (shaded region is excluded). The region in the black lines is the allowed $\epsilon_u$ obtain from the region between the solid black, dotted red, and dashed blue lines in Figs. \ref{fig:gp_exclusion} and \ref{fig:Rnp_exclusion} from our previous work \cite{Liu:2016qwd,Liu:2017bzj}. The vertical line indicates the allowed mass range obtained in Fig. \ref{fig:eta_to_pi_gamma_gamma}.
}
\end{figure}

\begin{figure}
\centering
\includegraphics[width=\columnwidth]{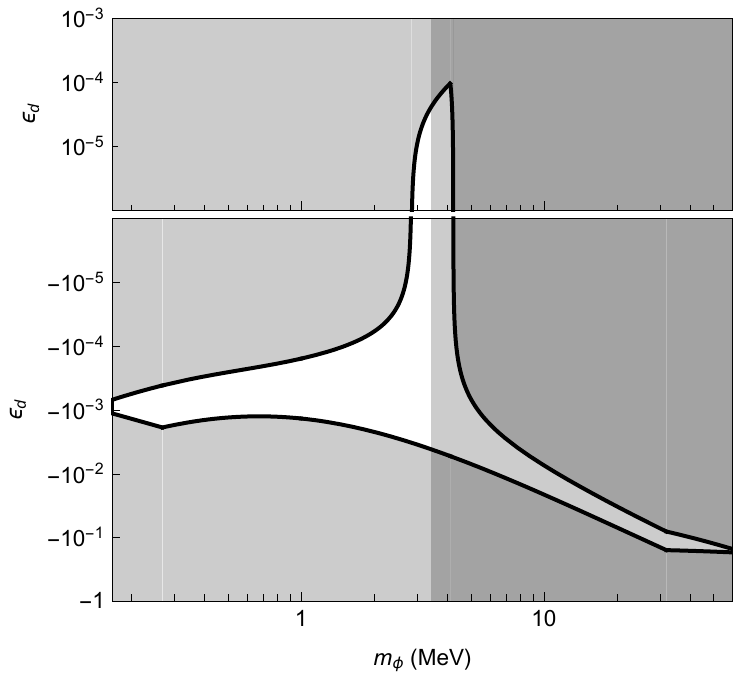}
\caption{\label{fig:gd_exclusion} Exclusion plot for $\epsilon_d$ (shaded region is excluded). The region in the black lines is the allowed $\epsilon_d$ obtain from the region between the solid black, dotted red, and dashed blue lines in Figs. \ref{fig:gp_exclusion} and \ref{fig:Rnp_exclusion} from our previous work \cite{Liu:2016qwd,Liu:2017bzj}. The vertical line indicates the allowed mass range obtained in Fig. \ref{fig:eta_to_pi_gamma_gamma}.
}
\end{figure}

Here we use a simplified Lagrangian including the new boson $\phi$ in the mostly plus metric:
\begin{align}
\mathcal{L}_\phi\supset-\frac{1}{2}(\partial\phi)^2-\frac{1}{2}m_\phi^2\phi^2+e\epsilon_f\phi\bar{\psi}_f\psi_f \label{L}
\end{align}
where $f$ is the flavor index, $\epsilon_f=g_f/e$, $e$ is the electric charge, and $\psi_f$ is the fermion field (quarks and leptons) in the SM.  The couplings to the neutron, $\epsilon_n$, and proton $\epsilon_p$  
are given by
\bea \epsilon_p=2\e_u+ \e_d,\quad\epsilon_n=2\e_d+ \e_u.
\eea
The Lamb shift in muonic hydrogen fixes $\epsilon_\mu$ and $\epsilon_p$ to have the same sign, therefore, we choose $\epsilon_\mu$ and $\epsilon_p$ to be positive, and $\epsilon_n$ and $\epsilon_e$ are allowed to have either sign. From our previous work \cite{Liu:2016qwd,Liu:2017bzj}, the allowed values of $\epsilon_p$ and $\epsilon_n$ are shown in Figs. \ref{fig:gp_exclusion} and \ref{fig:Rnp_exclusion} between the solid black, dotted red, and dashed blue lines. Since $\epsilon_n$ can take either sign, we present $\epsilon_n$ as a ratio to $\epsilon_p$. The allowed values of  $\epsilon_\mu$ are shown in Fig. \ref{fig:gmu_exclusion}. We can find the allowed regions of $\epsilon_u$ and $\epsilon_d$ in Fig. \ref{fig:gu_exclusion} and \ref{fig:gd_exclusion}, using $\epsilon_p$ and $\epsilon_n$ in Figs. \ref{fig:gp_exclusion} and \ref{fig:Rnp_exclusion}.

%===============================================================================
%===============================================================================

%===============================================================================
%===============================================================================

%%%%%%%%%%%%%%%%%%%%%%%%%%%%
\subsection{A Concrete UV Model}
Coupling a single scalar to $u$ and $d$ quarks in an effective Lagrangian at the MeV scale is not consistent with the $SU(2)_L \times U(1)_Y$ gauge invariance of the SM above the electroweak symmetry breaking scale. An explicit model  that specifies the $SU(2)_L \times U(1)_Y$ quantum numbers of new particles and the Lagrangian above and below the weak scale was constructed in the light ``Dark Higgs'' solution to the $(g-2)_\mu$ puzzle provided in Ref.~\cite{Chen:2015vqy}.
In their work, the low energy Lagrangian (that of our \eq{L}) appears as their low-energy theory. (We need only the CP-even dark Yukawa coupling here.) 

Ref.~\cite{Chen:2015vqy} provided a possible UV completion of the low energy effective theory of \eq{L}. In this framework, all new particles are assumed to be charged under $U(1)_d$ with the same dark charge so only their SM charges are identified.  They let $X^\ell$, where $\ell=e,\mu,\tau$ is a flavor index, be vector-like fermions with the quantum numbers of right-handed SM leptons
$\ell_R$ (i.e.~$SU(2)$ singlets), and masses $m_X^\ell \gsim \text{few}\times 100$~GeV. They also introduce a new Higgs scalar doublet $H_d$ and a complex scalar singlet $\phi$.  It is  assumed that $H_d$ and $\phi$ have nonzero vacuum expectation values (vevs) which spontaneously break $U(1)_d$. These ingredients can be motivated within a dark $Z$ model \cite{Davoudiasl:2012ag}.  Then they postulate the following ${\rm SM}\times U(1)_d$ invariant interactions
\begin{eqnarray}
-\mathcal{L}_1 &=& m_X^{\ell \ell'} \bar X^\ell X^{\ell'} + \lambda_1 \phi \bar X^\ell_L \ell_R	+ \lambda_2 H_d \bar L^\ell X^\ell_R \\
&+& y_\ell H \bar L^\ell \ell_R + {\rm h.c.}\nonumber
\end{eqnarray}
where $L^\ell$ and $H$ refer to SM lepton and Higgs doublets, respectively. The above interactions respect lepton flavor conservation up to soft breaking by (small) off-diagonal masses $m_X^{\ell \ell'}$, which are assumed to be the only sources of lepton flavor violation. Ref.~\cite{Chen:2015vqy}  illustrates how the above can be realized in a model with flavor symmetries that allow for a realistic neutrino mass matrix.  A vacuum expectation value for $H_d$ followed by charged lepton mass matrix diagonalization could result in misaligned $\phi$ and $H$ lepton couplings which lead to interesting consequences, as  discussed in their paper. Their model also includes scalar coupling to quarks. In that case the $H$ and $H_d$ alignment with the mass matrix is maintained and flavor changing current constraints are avoided at the tree level.

%%%%%%%%%%%%%%%%%%%%%%%%%%%%%

%===============================================================================
%===============================================================================
\section{The $\eta\pi^0\phi$ vertex}\label{sec:etapiphi}
%
%-------------------------------------------------------------------------------
\begin{figure}[tbp]
\centering
\includegraphics[width=\columnwidth]{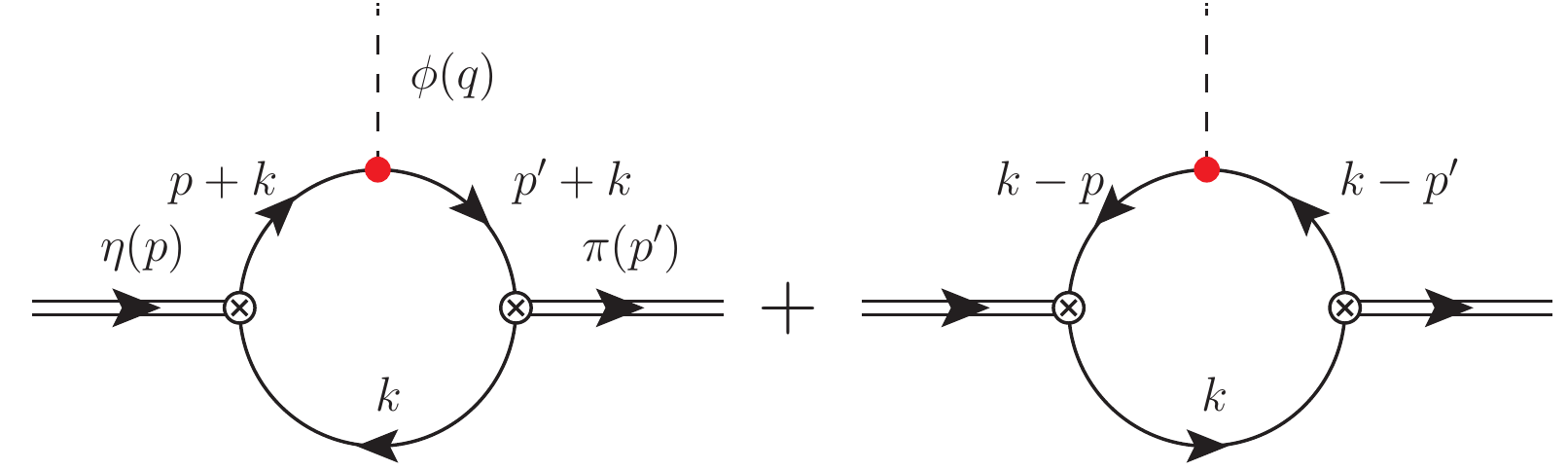}
\caption{Diagrams that contribute to the $\eta\to\pi^0\phi$ form factor. The single line is the dressed-quark propagator, the solid-dot is the $\phi$-quark vertex, and the crossed-circle is the appropriate Bethe-Salpeter vertex.}
\label{fig:eta_to_pi_phi}
\end{figure}
%-------------------------------------------------------------------------------

%
To use $\eta$  decay to constrain $\phi$ we need to know the $\eta\pi^0\phi$ vertex: $g'\eta\pi^0\phi$ to compute the $\eta$ decay rate and constrain the possible values of $\e_{u,d}$. To obtain the relevant vertex, the  $\pi^0$ and $\eta$  quark-model wave functions are used
\begin{align}
\pi^0=\frac{u\bar{u}-d\bar{d}}{\sqrt{2}}\quad\text{and}\quad\eta=\frac{u\bar{u}+d\bar{d}-2s\bar{s}}{\sqrt{6}},
\end{align}
the coupling $g'$ reads
\begin{align}
g'=2f\left(\frac{g_u}{\sqrt{2}\sqrt{6}}-\frac{g_d}{\sqrt{2}\sqrt{6}}\right)=f\ \frac{g_u-g_d}{\sqrt{3}}
\end{align}
where the overall factor of 2 takes into account both the $q$ and $\bar{q}$ contributions;  the dimension of $f$ is that of mass. Recall that $\e_f=g_f/e$.

Since $\phi$ actually couples to the quarks, the $\eta\pi^0\phi$ vertex should be described by a form factor  that accounts for   meson structure.  We use the Nambu--Jona-Lasinio (NJL) model~\cite{Nambu:1961fr,Nambu:1961tp,Vogl:1991qt,Klevansky:1992qe} to estimate this form factor. The piece of the three-flavor NJL Lagrangian relevant to this problem reads
\begin{align}
\mathcal{L}_{NJL}\supset \bar{\psi}(i\slashed{\partial}-\hat{m})\psi + G_\pi\left[(\bar{\psi}\lambda_a\psi)^2-(\bar{\psi}\lambda_a\gamma_5\psi)^2\right],
\end{align}
where $\psi = (\psi_u,\psi_d,\psi_s)$ is the quark field, $\hat{m}={\rm diag}(m,m,m_s)$ is the  current quark mass matrix (here we have set $m_u=m_d=m$), and $\lambda_a$ are the Gell-Mann matrices with $\lambda_0=\sqrt{\frac{2}{3}}\,\mathbb{1}$. The $\eta\pi^0\phi$ form factor in Fig. \ref{fig:eta_to_pi_phi} is found to be
\begin{align}
g'(p,p') &= N_c\,i\!\int\!\frac{d^4k}{(2\pi)^4} \, {\rm Tr}\bigg[\Gamma_\pi\hat{S}(p'+k)\,\hat{g}\,\hat{S}(p+k)\Gamma_\eta\,\hat{S}(k) \nonumber\\
&\hspace*{18mm} + \Gamma_\eta\hat{S}(k-p)\hat{g}\hat{S}(k-p')\Gamma_\eta\hat{S}(k)\bigg],
\label{eq:formfactor}
\end{align}
where $N_c$ is number of colors, the $\pi^0$ and $\eta$ Bethe-Salpeter vertices~\cite{Cloet:2014rja} are given by $\Gamma_\pi=\sqrt{Z_\pi}\,\gamma_5\lambda_3$ and $\Gamma_\eta=\sqrt{Z_\eta}\,\gamma_5\lambda_8$\footnote{In this calculation we ignore $\eta-\eta'$ mixing.}, $\hat{g}={\rm diag}(g_u,g_d,g_s)$, and $\hat{S}={\rm diag}(S,\,S,\,S_s)$ with 
\begin{align}
S(p) = \frac{1}{-\sh{p} - M + i\ve} \quad {\rm and} \quad S_s(p) =  \frac{1}{-\sh{p} - M_s + i\ve},
\end{align}
where $M,M_s$ are the dressed quark masses. These are generated by the spontaneous symmetry breaking famously (Nobel Prize 2008)   inherent in the NJL model. Performing the flavor space trace in Eq.~\eqref{eq:formfactor}, and using the proper-time regularization scheme~\cite{Schwinger:1951nm,Ebert:1996vx,Hellstern:1997nv,Cloet:2014rja} gives
\begin{align}
g'(p',p)&= \frac{g_u-g_d}{\sqrt{3}}\ f(q^2),
\end{align}
where $q = p'-p$ and
\begin{align}
& f(q^2)=\frac{3M}{2\pi^2}\,\sqrt{Z_\eta Z_\pi}\int_{\tau_{ir}}^{\tau_{uv}}\!\! d\tau \int_0^1 dx\, \bigg\{\frac{1}{\tau}\,e^{-\tau\lf[M^2 + x(1-x)q^2\rg]} \no  \\ 
&+\frac{m_\eta^2+m_\pi^2+q^2}{2}\int_0^{1-x}\!\!dy\, e^{-\tau\lf[(x+y-1)(x\,m_\pi^2+y\,m_\eta^2) + M^2 + xyq^2\rg]}\bigg\}.
\end{align}
The integral over proper-time  includes  both an infrared and ultraviolet cutoff, namely, $\tau_{ir} = 1/\Lambda_{UV}^2$ and $\tau_{uv} = 1/\Lambda_{IR}^2$, where the former implements aspects of quark confinement~\cite{Cloet:2014rja}. 
%The functions in the exponential take the form $h_1(x) = M^2 + x(1-x)q^2$ and $h_2(x,y) = (x+y-1)(x\,m_\pi^2+y\,m_\eta^2) + M^2 + xyq^2$. 
Since we are interested in the decay rate of $\eta\to\pi^0\gamma\gamma$ or $\eta\to\pi^0\bar{\psi}_f\psi_f$ process, the allowed values of  $q^2$ range from 0 to $-(m_\eta-m_\pi)^2$ if $\eta$ is at rest (the lower bound of $q^2$ is exact for $\pi^0\gamma\gamma$ final state and a good approximation for $\pi^0\bar{\psi}_f\psi_f$ final state with light fermions).

To constrain $f(q^2)$ we consider dressed quark masses in the range $200\,{\rm MeV} < M < 400\,{\rm MeV}$, with values outside this range deemed unlikely based on previous studies of meson and nucleon properties~\cite{Ninomiya:2014kja,Cloet:2014rja}. For each dressed-quark mass the parameters $\Lambda_{UV}$ and $G_\pi$ are adjusted so that the empirical values of the pion mass and decay constant are reproduced, and $\Lambda_{IR}$ should be of the order of $\Lambda_{\rm QCD}$ because it represents the effects of confinement in the model, and we chose $\Lambda_{IR} = 240\,$MeV. Our results are summarized in  Table~\ref{table:NJL parameters}, where the Bethe-Salpeter vertex normalizations, $Z_\eta$ and $Z_\pi$, are outputs of the calculation.

The range of values for $f(q^2)$ are summarized in Table~\ref{table:NJL parameters}. The form factor $f(q^2)$ can take values from 551 to 1274\,MeV including momentum dependence and model dependence. Within our range of interest, we can drop the momentum dependence and treat $f$ as a coupling constant, with a magnitude and uncertainty given by
\begin{align}
f=825^{+450}_{-275}\ {\rm MeV}
\label{eq:f}
\end{align}
where the central value of $f$ is chosen to be 825 MeV, such that the associate parameters give the best description of pion and kaon system~\cite{Ninomiya:2014kja}. We will use the lower bound of $f$ to constrain the scalar boson $\phi$.

One can also consider $\eta-\eta'$ mixing~\cite{Cao:2012nj,Bramon:1997va,Pham:2015ina,Christ:2010dd,Ottnad:2012fv}:
\begin{align}
\left(\begin{array}{c} \eta \\ \eta' \end{array}\right)=
\left({\begin{array}{cc} \cos\theta & -\sin\theta \\ \sin\theta & \cos\theta \end{array}}\right)\left(\begin{array}{c} \eta_8 \\ \eta_1 \end{array}\right) .
\end{align}
The influence of $\eta-\eta'$ mixing implies the following replacement in the previous equations:
\begin{align}
\eta=\sqrt{3}\cos(\theta+\tan^{-1}\sqrt{2})\eta_8,
\end{align}
where $\theta$ ranges from $-17^\circ$ to $-12^\circ$ which corresponds to approximately a 30\% change in $f(q^2)$. However, $\eta-\eta'$ mixing is also model dependent in our calculation, therefore we will include the effect of $\eta-\eta'$ mixing   in the estimate of the model-dependent uncertainties already given in Eq.~\eqref{eq:f}.

\begin{table}[tbp]
\centering
\addtolength{\tabcolsep}{7.0pt}
\addtolength{\extrarowheight}{2.2pt}
\begin{tabular}{cccccc} 
\hline\hline
%$M$ (MeV) & $\Lambda_{UV}$ (MeV) & $G_\pi$ (GeV$^{-2}$) & $Z_\eta$ & $Z_\pi$ & $f(q^2)$ (MeV) \\
$M$ & $\Lambda_{UV}$ & $G_\pi$ & $Z_\eta$ & $Z_\pi$ & $f(q^2)$ \\
\hline
200 & 1282 & 2.209 & 4.603 & 4.922 & 551$\sim$577 \\ 
300 & 715 & 10.38 & 10.27 & 11.52 & 909$\sim$962 \\ 
400 & 638 & 19.84 & 18.28 & 20.86 & 1205$\sim$1274 \\ 
\hline\hline
\end{tabular}
\caption{\label{table:NJL parameters} Results and parameters in NJL model for different dressed quark masses. The range of $f(q^2)$ corresponds to $q^2$ from 0 to $-(m_\eta-m_\pi)^2$. Dimensionful quantities are in units of MeV with the exception of $G_\pi$ which is in units of GeV$^{-2}$. }
\end{table}
%-------------------------------------------------------------------------------

%===============================================================================
%===============================================================================
\section{decay rate}\label{sec:decay rate}

\subsection{$\phi$ decay rate}\label{sec:phi decay rate}

\begin{figure}[tbp]
\centering
\includegraphics[width=\columnwidth]{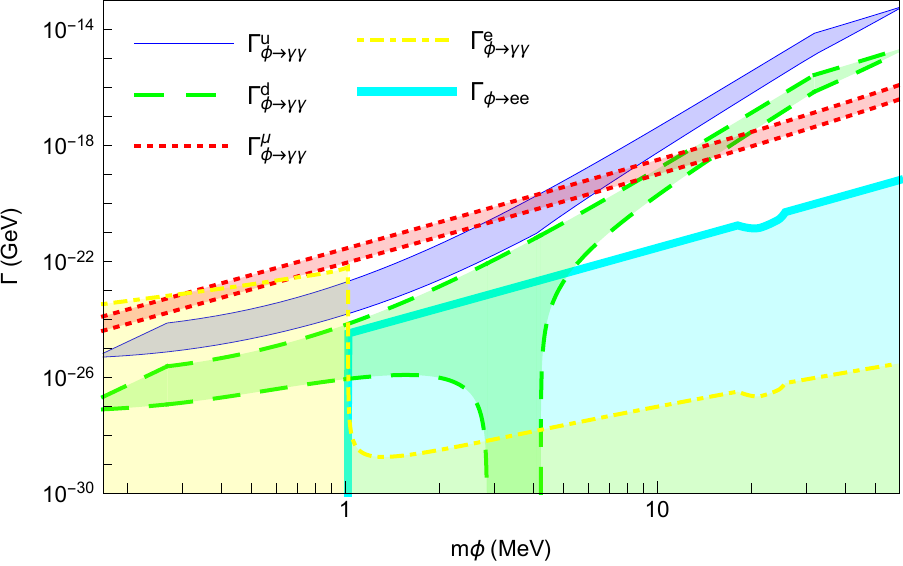}
\caption{\label{fig:phi_decay_rate} The total $\phi$ decay width (shaded region is allowed to decay). The decay width with a superscript $f$ indicates the process decaying through $f$ fermion loop. The thin solid blue, dashed green, and dotted red lines are $\phi\to\gamma\gamma$ through $u$ quark, $d$ quark, and muon loops with couplings from Figs. \ref{fig:gu_exclusion}, \ref{fig:gd_exclusion}, and \ref{fig:gmu_exclusion}, respectively. The thick solid cyan and dotted dashed yellow lines are $\phi\to e^+e^-$ and $\phi\to\gamma\gamma$ through electron loop with the coupling $\epsilon_e$ from Fig. \ref{fig:ge_exclusion}.
}
\end{figure}

If $m_\phi>2m_e$, $\phi$ can decay to two fermions and the decay width is
\begin{align}\label{eq:phi to pi f f}
\Gamma_{\phi\to f\bar{f}}=\epsilon_f^2\,\frac{\alpha}{2}\,m_\phi\left(1-\frac{4m_f^2}{m_\phi^2}\right)^{3/2},
\end{align}
where $\alpha$ is the fine structure constant and $m_f$ is the fermion mass. If $\phi$ decays to two photons through a fermion loop
\begin{align}\label{eq:phi to gamma gamma}
\Gamma_{\phi\to\gamma\gamma}^f=\epsilon_f^2\,Q_f^4\,\frac{\alpha^3}{4\pi^2}\,\frac{m_\phi^3}{m_f^2}\left|I\left(\frac{4m_f^2}{m_\phi^2}\right)\right|^2,
\end{align}
where the superscript $f$ of $\Gamma$ indicates the fermion in the loop; $Q_f$ is the electric charge of the fermion in the units of $e$, e.g. $Q_u=2/3$; $I$ is obtained in Ref.~\cite{Gunion:1989we} and reads
\begin{align}\label{eq:I(tau)}
I(\tau) &= \int_0^1dx_1\int_0^{1-x_1}dx_2\frac{1-4x_1x_2}{1-\frac{4}{\tau}x_1x_2}, \nonumber \\
&=\frac{\tau}{2}[1+(1-\tau)f(\tau)]
\end{align}
where
\begin{align}
f(\tau)=
\begin{cases}
    \left(\sin^{-1}\frac{1}{\sqrt{\tau}}\right)^2,& \text{if } \tau\geq 1,\\
    -\frac{1}{4}\left[\ln\left(\frac{1+\sqrt{1-\tau}}{1-\sqrt{1-\tau}}\right)-i\pi\right]^2,& \text{if } \tau<1.
\end{cases}
\end{align}
The total $\phi$ decay rate is
\begin{align}
\Gamma_{\phi,{\rm total}} &= \Gamma_{\phi\to e^+e^-}\theta(m_\phi-2m_e)+\Gamma_{\phi\to\gamma\gamma}^e\no \\
&\hs*{7mm}
+\Gamma_{\phi\to\gamma\gamma}^\mu+\Gamma_{\phi\to\gamma\gamma}^u
+\Gamma_{\phi\to\gamma\gamma}^d+\Gamma_{\phi\to\gamma\gamma}^\text{interference},
\end{align}
where $\Gamma_{\phi\to\gamma\gamma}^\text{interference}$ is the interference of different fermions loops. We use the constituent quark mass ($m_u=m_d=200$ MeV
%~\cite{Tanabashi:2018oca}
) for the decay through quark loop, because the relevant scale of the $\phi$ decay process is $m_\phi$ and the quark mass should accordingly be evolved to this scale.
The result is shown in Fig.~\ref{fig:phi_decay_rate}. Since the interference contribution is expected to be smaller than the leading one, we neglect $\Gamma_{\phi\to\gamma\gamma}^\text{interference}$. 

\subsection{$\eta$ decay rate}
The total decay width of $\eta$, $\Gamma_{\eta,\text{total}}$, is 1.31$\pm$0.05 keV \cite{Patrignani:2016xqp}. Since the new scalar particle $\phi$ decays into two fermions or two photons final state,  $\eta$ decay may include  the process
\begin{align}
\eta(p)\to\pi^0(p_2)+\phi^*(p_1+p_3)\to\pi^0(p_2)+X(p_1)+\overline{X}(p_3)
\end{align}
where $X$ can be either fermion or photon. The three body final state phase space integral is \cite{Asatrian:2012tp}
\begin{align}
D\Phi_3=\frac{m_\eta^2}{64\pi^3}\tilde{p}\,\tilde{p}_3\, d\cos\theta ds_{13}
\end{align}
where $\tilde{p}$, $\tilde{p}_3$, and $s_{13}$ are dimensionless and given by
\begin{align}
\tilde{p}&=\frac{\sqrt{(1+x_2-s_{13})^2-4x_2}}{2\sqrt{s_{13}}},\\
\tilde{p}_3&=\frac{\sqrt{(x_1+x_3-s_{13})^2-4x_1x_3}}{2\sqrt{s_{13}}},\\
s_{13}&=\frac{-(p_1+p_3)^2}{m_\eta^2}=\frac{2m_X^2-2p_1\cdot p_3}{m_\eta^2};
\end{align}
$x_2=m_{\pi^0}^2/m_\eta^2$ and $x_1=x_3=m_X^2/m_\eta^2$;
$\cos\theta$ is the polar angle of $\mathbf{p}_3$ with respect to $\mathbf{p}$. Therefore, the three body final state decay rate is 
\begin{align}\label{eq:three body decay rate}
\Gamma &=\frac{1}{S}\int\frac{1}{2m_\eta}\overline{|\mathcal{M}|^2}D\Phi_3, \no \\
&=\frac{m_\eta}{128\pi^3 S}\int_{-1}^1 d\cos\theta \int_{s_{13}^{min}}^{s_{13}^{max}} ds_{13} \overline{|\mathcal{M}|^2}\tilde{p}\,\tilde{p}_3
\end{align}
where $S$ is the symmetry factor taking into account how many identical particles in the final state, ${s_{13}^{min}}=(\sqrt{x_1}+\sqrt{x_3})^2$, and ${s_{13}^{max}}=(1-\sqrt{x_2})^2$.

\section{beam dump experiments}\label{sec:beam dump experiments}
\begin{figure}
\centering
\includegraphics[width=\columnwidth]{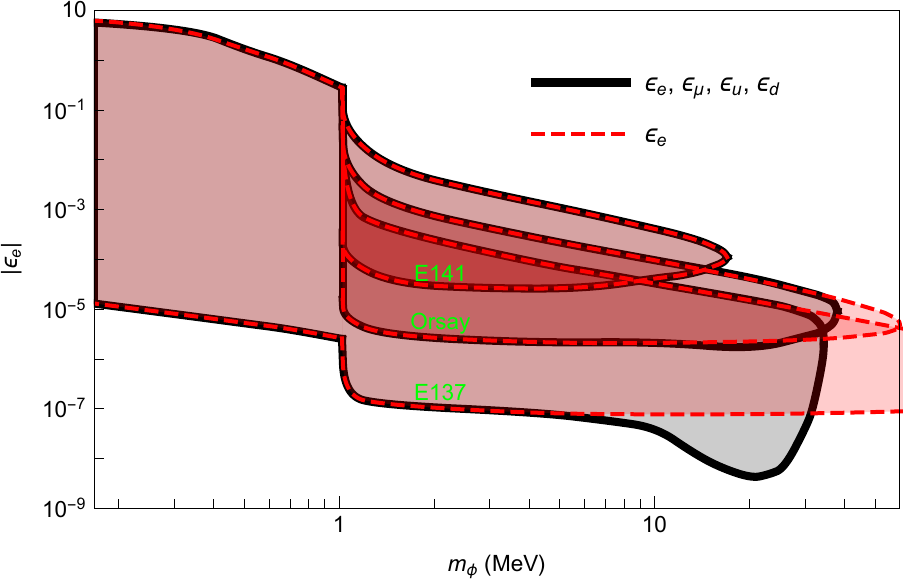}
\caption{\label{fig:BD} The constraint of beam dump experiments (shaded regions are excluded). The region in the dashed red line is obtained if $\phi$ only couples to an electron. The region in the black line includes the $\phi$ coupling to a muon, $u$ quark, and $d$ quark.}
\end{figure}

In our previous work \cite{Liu:2016qwd,Liu:2017bzj}, we considered the constraints of beam dump experiments \cite{Liu:2016mqv,Bjorken:1988as,Riordan:1987aw,Davier:1989wz}. However, in making the  $\epsilon_e$ exclusion plots we only included $\phi\to 2e$, $\phi\to 2\mu$, and $\phi\to 2\gamma$ as proceeding through the electron loop. Since $\epsilon_\mu$, $\epsilon_u$ and $\epsilon_d$ are much larger than $\epsilon_e$, we should include $\phi\to 2\gamma$ through muon, $u$ quark, and $d$ quark loops as well. Further investigation and recalculation shows that the exclusion plots change quite a lot, see Fig. \ref{fig:BD}. 

The changes of the exclusion plots are easy to explain. The coupling of $\epsilon_u$ and $\epsilon_d$ become bigger in the large mass region as well as the decay width of $\phi\to 2\gamma$ through quark loops. The decay length of $\phi$ become shorter than the thickness of the shield so the exclusion stops when $m_\phi\gtrsim 30$ MeV. On the other hand, $\Gamma^u_{\phi\to\gamma\gamma}$ becomes dominant when $m_\phi>10$ MeV. The constraint for $\epsilon_e$ need to be smaller so that $\phi$ is harder to produce from beam dump. Therefore the lower bound of the constraint is lower. The result is also shown in Fig. \ref{fig:ge_exclusion}.

\section{$\eta$ decay to two leptons ($\eta\to\pi^0\phi^*\to\pi^0 f\bar{f}$)}\label{sec:eta to pi f f}
We emphasize that if the intermediate state $\phi$ is a real particle, its decay products are extremely difficult to detect since $\phi$ is long-lived. Therefore we consider the virtual scalar boson decay to two fermion channel. The amplitude is
\begin{align}
\mathcal{M}=\frac{g'g_f}{(p_1+p_3)^2+m_\phi^2}\,\bar{u}_1 v_3
\end{align}
and  summing the square of the amplitude over final states gives
\begin{align}
\overline{|\mathcal{M}|^2}=\epsilon_f^2(\epsilon_u-\epsilon_d)^2f^2\frac{32\pi^2\alpha^2}{3} \frac{s_{13}m_\eta^2-4m_f^2}{(s_{13}m_\eta^2-m_\phi^2)^2},
\end{align}
where $s_{13}m_\eta^2=-(p_1+p_3)^2=2m_f^2-2p_1\cdot p_3$. Using Eq. (\ref{eq:three body decay rate}), the decay rate is found to be
\begin{align}\label{eq:eta to pi f f}
\Gamma(\eta\to\pi\bar{\psi}_f\psi_f) &= \epsilon_f^2(\epsilon_u-\epsilon_d)^2\alpha^2\frac{f^2m_\eta}{6\pi} \no \\
&\hs*{4mm}
\int_{s_{13}^{min}}^{s_{13}^{max}} ds_{13} \frac{s_{13}m_\eta^2-4m_f^2}{(s_{13}m_\eta^2-m_\phi^2)^2}\tilde{p}\,\tilde{p}_3.
\end{align}

\subsection{$\eta\to\pi^0\mu^+\mu^-$}
\begin{figure}[tbp]
\centering
\includegraphics[width=\columnwidth]{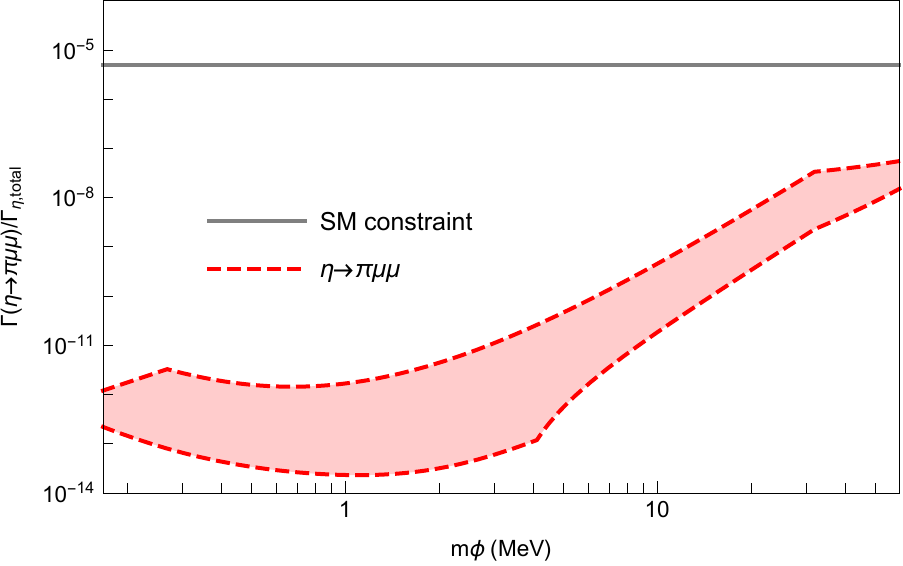}
\caption{\label{fig:eta_to_pi_mu_mu} Decay rate of $\eta\to\pi^0\phi^*\to\pi^0\mu^+\mu^-$ (shaded region is allowed to decay). Since the decay rate of this channel is much smaller than the SM constraint, there is no new constraint obtained from this channel.}
\end{figure}

 The process $\eta\to\pi^0 \mu^+\mu^-$, which  involves the decay of the virtual $\phi$ ($\phi^*\to\mu^+\mu^-$) has not been observed. The present constraint is 
\begin{align}
\frac{\Gamma(\eta\to\pi^0 \mu^+\mu^-)}{\Gamma_{\eta,\text{total}}}<5\times 10^{-6}\quad\text{at CL = 90\%}.
\end{align}
In Fig. \ref{fig:eta_to_pi_mu_mu}, using Eq. (\ref{eq:eta to pi f f}) we show that the new channel to muon pair through a virtual $\phi$ is much smaller than the the SM constraint, therefore no new constraint is obtained from this channel.

\begin{figure}
\centering
\includegraphics[width=\columnwidth]{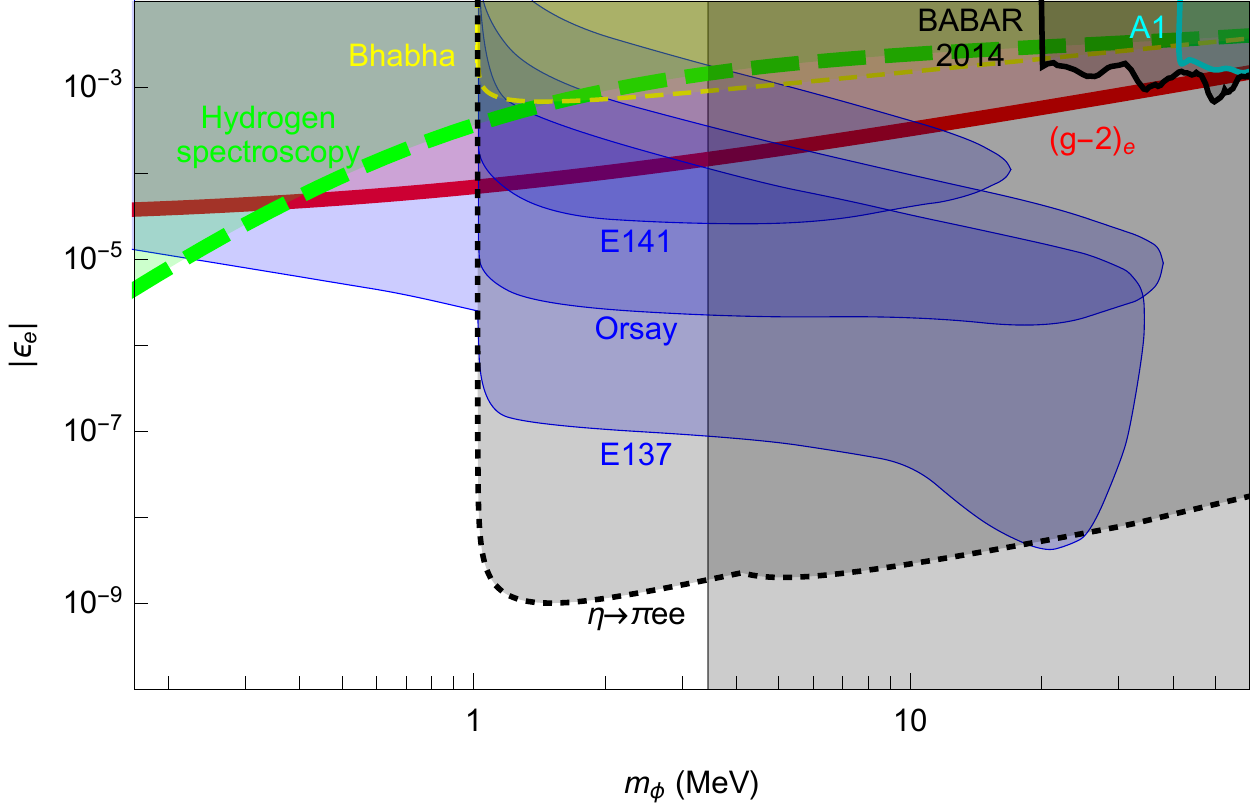}
\caption{\label{fig:ge_exclusion} Exclusion plot for $|\epsilon_e|$ (shaded region is excluded). The thin blue, thick red, thin dashed yellow, and thick dashed green lines are from our previous work \cite{Liu:2016qwd,Liu:2017bzj} corresponding to beam dump experiments (see, Fig. \ref{fig:BD}), electron anomalous magnetic moment $(g-2)_e$ \cite{Pospelov:2008zw,Bouchendira:2010es}, Bhabha scattering \cite{Tsertos:1989gv}, and the Lamb shift of hydrogen \cite{Miller:2011yw,Miller:2012ht,Miller:2012ne,Eides:2000xc}. A1 at MAMI \cite{{Merkel:2014avp}} and BABAR 2014 \cite{{Lees:2014xha}} constraints are in the upper right corner. The dotted black line is our new result from $\eta\to\pi^0e^+e^-$ decay. The vertical line indicates the allowed mass range obtained in Fig. \ref{fig:eta_to_pi_gamma_gamma}.
}
\end{figure}

\subsection{$\eta\to\pi^0 e^+e^-$}
The process $\eta\to\pi^0 e^+e^-$ has not been observed, and the constraint is 
\begin{align}
\frac{\Gamma(\eta\to\pi^0 e^+e^-)}{\Gamma_{\eta,\text{total}}}<4\times 10^{-5}\quad\text{at CL = 90\%}.
\end{align}
The decay of the virtual $\phi$ to electron-positron pairs,  $\phi^*\to e^+e^-$ would contribute to this rate.

We have to handle this process with more care. If $m_\phi>2m_e$, the virtual $\phi$ propagator can be on-shell and we need to put in the total $\phi$ decay width
\begin{align}
(s_{13}m_\eta^2-m_\phi^2)^2\to(s_{13}m_\eta^2-m_\phi^2)^2+m_\phi^2\Gamma_{\phi,\text{total}}^2.
\end{align}
We can further use the narrow width approximation (NWA)
\begin{align}
\frac{1}{(s-m^2)^2+m^2\Gamma^2}\to\frac{\pi}{m\Gamma}\delta(s-m^2) \quad\text{if}\;\;\frac{\Gamma}{m}\to 0.
\end{align}
Assuming $m_\phi > 2\,m_e$ the decay rate becomes
\begin{multline}\label{eq:eta to pi e e with NWA}
\Gamma(\eta\to\pi ee)_\text{NWA} \\ 
=\epsilon_e^2(\epsilon_u-\epsilon_d)^2\alpha^2 \left. \frac{f^2(m_\phi^2-4m_e^2)}{6m_\eta m_\phi\Gamma_{\phi,\text{total}}} \tilde{p}\,\tilde{p}_3\right|_{s_{13}=\frac{m_\phi^2}{m_\eta^2}}.
\end{multline}
Using Eqs.~\eqref{eq:eta to pi f f} and \eqref{eq:eta to pi e e with NWA}, we show the exclusion of $\epsilon_e$ in Fig. \ref{fig:ge_exclusion}.

%===============================================================================
%===============================================================================
\section{Eta decay to two photons ($\eta\to\pi^0\phi^*\to\pi^0 \gamma\gamma$)}\label{sec:eta to pi gamma gamma}
Based on the same reason argued in Sec.~\ref{sec:eta to pi f f}, we only consider the intermediate state $\phi$ is virtual. The amplitude $\eta\to\pi^0 \gamma\gamma$ through a virtual $\phi$ and then a fermion loop with flavor $f$ (two diagrams) is
\begin{align}
&i\mathcal{M} = ig'\frac{-i}{(p_1+p_3)^2+m_\phi^2} \no \\
&\hs*{1mm}
\left[
-iQ_f^2\frac{2g_f\alpha}{\pi m_f}\epsilon_\mu^{\lambda_1}\epsilon_\nu^{\lambda_3}(-p_1\cdot p_3 g^{\mu\nu}+p_3^\mu p_1^\nu)I\left(\frac{4m_f^2}{s_{13}m_\eta^2}\right)\right],
\end{align}
where $Q_f$ is the fermion electric charge in units of $e$, e.g. $Q_d=-1/3$; $\epsilon_\mu^{\lambda}$ is the photon polarization vector; the term in the square bracket includes the contribution of two Feynman diagrams; $I$ is defined in Eq.~\eqref{eq:I(tau)}.

The amplitude squared (assuming there is only one fermion loop) after summing  over final states is
\begin{align}
\overline{|\mathcal{M}|^2}=\epsilon_f^2(\epsilon_u-\epsilon_d)^2\alpha^4Q_f^4\frac{32f^2}{3m_f^2} \frac{s_{13}^2m_\eta^4}{(s_{13}m_\eta^2-m_\phi^2)^2}\left|I\left(\frac{4m_f^2}{s_{13}m_\eta^2}\right)\right|^2.
\end{align}
Using Eq.~\eqref{eq:three body decay rate}, the decay rate is found to be 
\begin{align}
\Gamma&=\epsilon_f^2(\epsilon_u-\epsilon_d)^2\alpha^4 Q_f^4\frac{f^2m_\eta}{12\pi^3 m_f^2} \no \\
&\hs*{10mm}
\int_{0}^{s_{13}^{max}} ds_{13} \frac{s_{13}^2m_\eta^4}{(s_{13}m_\eta^2-m_\phi^2)^2} \left|I\left(\frac{4m_f^2}{s_{13}m_\eta^2}\right)\right|^2\tilde{p}\,\tilde{p}_3,\no \\
&=\epsilon_f^2(\epsilon_u-\epsilon_d)^2\alpha^4 Q_f^4\frac{f^2m_\eta^5}{24\pi^3 m_f^2} \no \\
&\hs*{10mm}
\int_{0}^{s_{13}^{max}} ds_{13} \frac{s_{13}^{5/2}}{(s_{13}m_\eta^2-m_\phi^2)^2} \left|I\left(\frac{4m_f^2}{s_{13}m_\eta^2}\right)\right|^2\tilde{p}.
\end{align}
We used the fact that there are two photons in the final state ($S=2$), $s_{13}^{min}=0$, and $\tilde{p}_3=\sqrt{s_{13}}/2$. We can further apply the narrow width approximation
\begin{multline}\label{eq:eta to pi gamma gamma with NWA}
\Gamma = \epsilon_f^2(\epsilon_u-\epsilon_d)^2\alpha^4 Q_f^4 \\ 
\times
\frac{f^2 m_\phi^4}{24\pi^2  m_f^2m_\eta^2\Gamma_{\phi,\text{total}}}\left|I\left(\frac{4m_f^2}{m_\phi^2}\right)\right|^2\tilde{p}\,\bigg|_{s_{13}=\frac{m_\phi^2}{m_\eta^2}}.
\end{multline}
The process $\eta\to\pi^0 \gamma\gamma$ is observed and the value is 
\begin{align}
\frac{\Gamma(\eta\to\pi^0 \gamma\gamma)}{\Gamma_{\eta,\text{total}}}=(2.56\pm 0.22)\times10^{-4}.
\end{align}
The decay rate of a virtual $\phi$ to two photons cannot be too big to spoil the observed value. We define that the scalar boson is excluded if the decay rate is greater than the observed value plus $3\sigma$, \textit{i.e.} $\Gamma(\eta\to\pi^0 \gamma\gamma)/\Gamma_{\eta,\text{total}}>3.22\times10^{-4}$.

\begin{figure}[tbp]
\centering\includegraphics[width=\columnwidth]{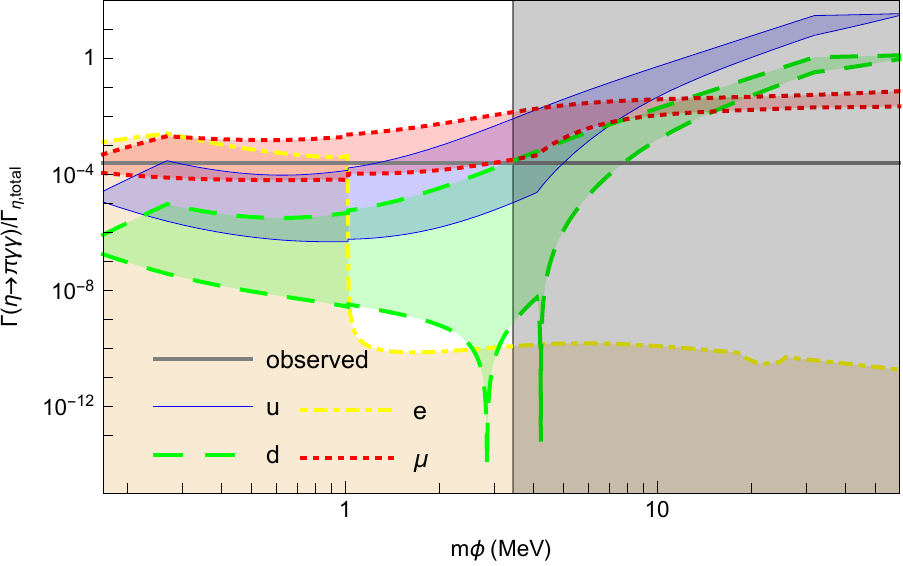}
\caption{\label{fig:eta_to_pi_gamma_gamma} Decay rate of $\eta\to\pi^0\phi^*\to\pi^0\gamma\gamma$ (shaded region is allowed to decay). The decay width with a superscript $f$ indicates the process decaying through $f$ fermion loop. The solid blue, dashed green, dotted red, and dotted dashed yellow lines are $\eta\to\phi\gamma\gamma$ through $u$ quark, $d$ quark, muon, and electron loops. The horizontal gray line is the observed decay width. The vertical line indicates where the decay rate of $\eta\to\pi\gamma\gamma$ through muon loop channel is greater than observed value plus $3\sigma$.
}
\end{figure}
In Fig. \ref{fig:eta_to_pi_gamma_gamma}, we show a virtual $\phi$ decay to two photons through different fermion loops. We can read from the plot that the allowed scalar boson mass is from 168 keV (from our previous work) to 3.45 MeV (from muon loop decay channel). Again, we neglect the interference terms because they are expected to be smaller than the leading term. 
The allowed $\epsilon_u$, $\epsilon_p$, and $\epsilon_\mu$ are all around $10^{-3}$.
One might think that the existence of the scalar boson could survive the constraints if its coupling to the $u$ quark (or muon) were zero. However, investigating the allowed regions of the parameter spaces, we find that  this is not the case. The exclusion of $\phi$ coupling to $u$ quark (muon) is equivalent to excluding the  existence of the $\phi$. 

%===============================================================================
%===============================================================================
\section{Coexistence with the Old and the New}\label{sec:coexistence}
\begin{figure}[!ht]
\centering
\includegraphics[width=\columnwidth]{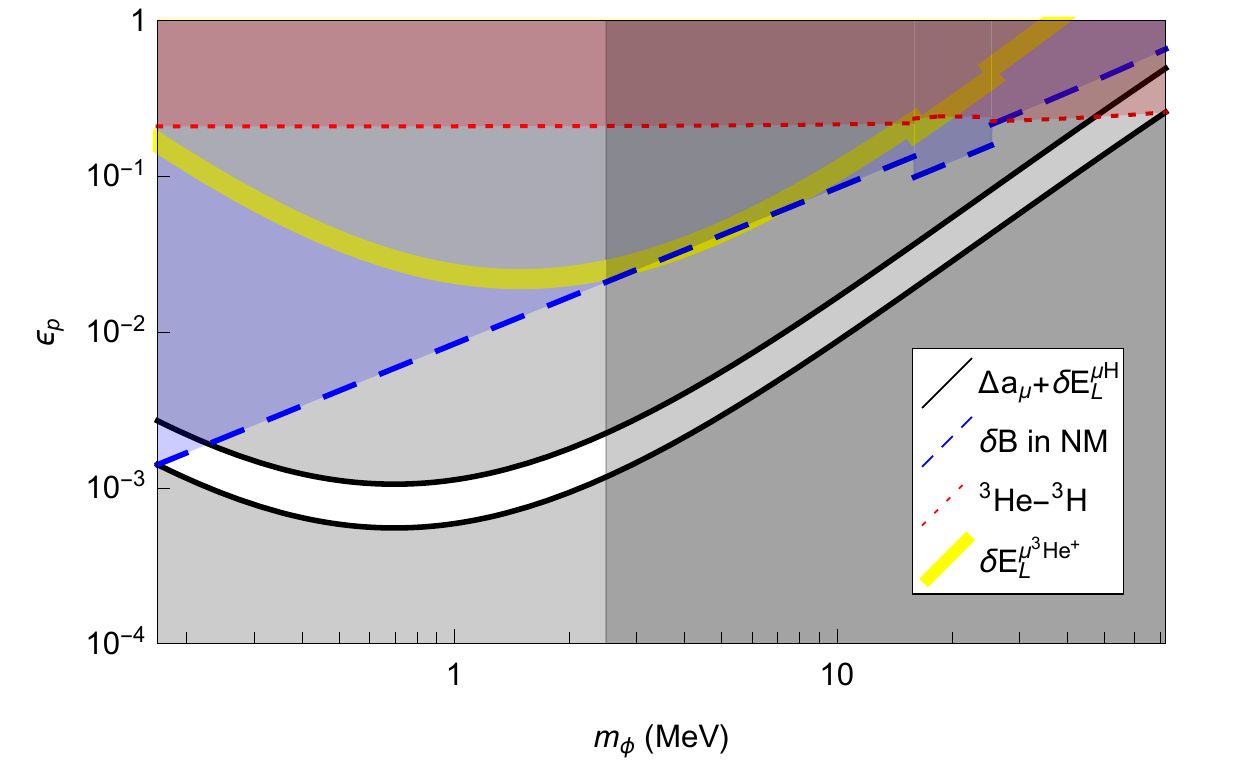}
\caption{\label{fig:gp_exclusion+newH} Exclusion plot for $\epsilon_p$ (shaded region is excluded). See caption of Fig.~\ref{fig:gp_exclusion} for more details.
}
\end{figure}

\begin{figure}[!ht]
\centering
\includegraphics[width=\columnwidth]{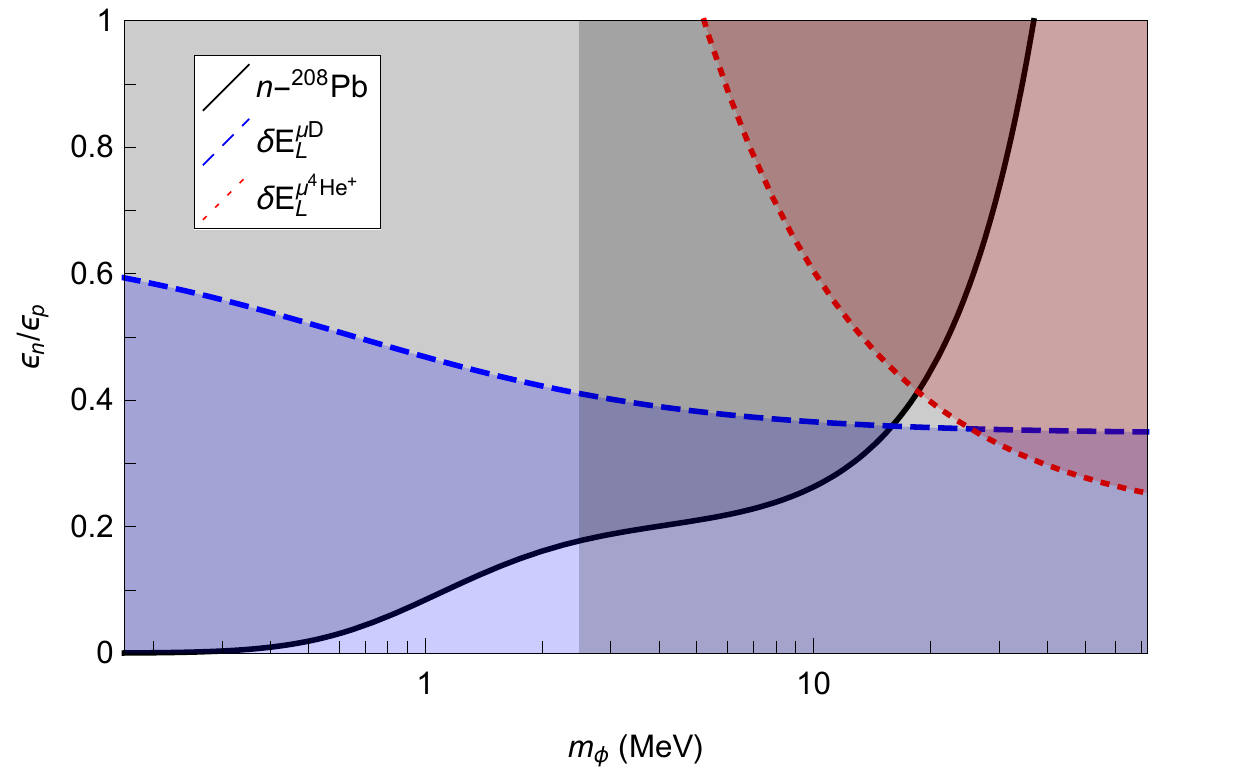}
\caption{\label{fig:Rnp_exclusion+newH} Exclusion plot for $\epsilon_n$ (shaded region is excluded). See caption of Fig.~\ref{fig:Rnp_exclusion} for more details.
}
\end{figure}

\begin{figure}[!ht]
\centering
\includegraphics[width=\columnwidth]{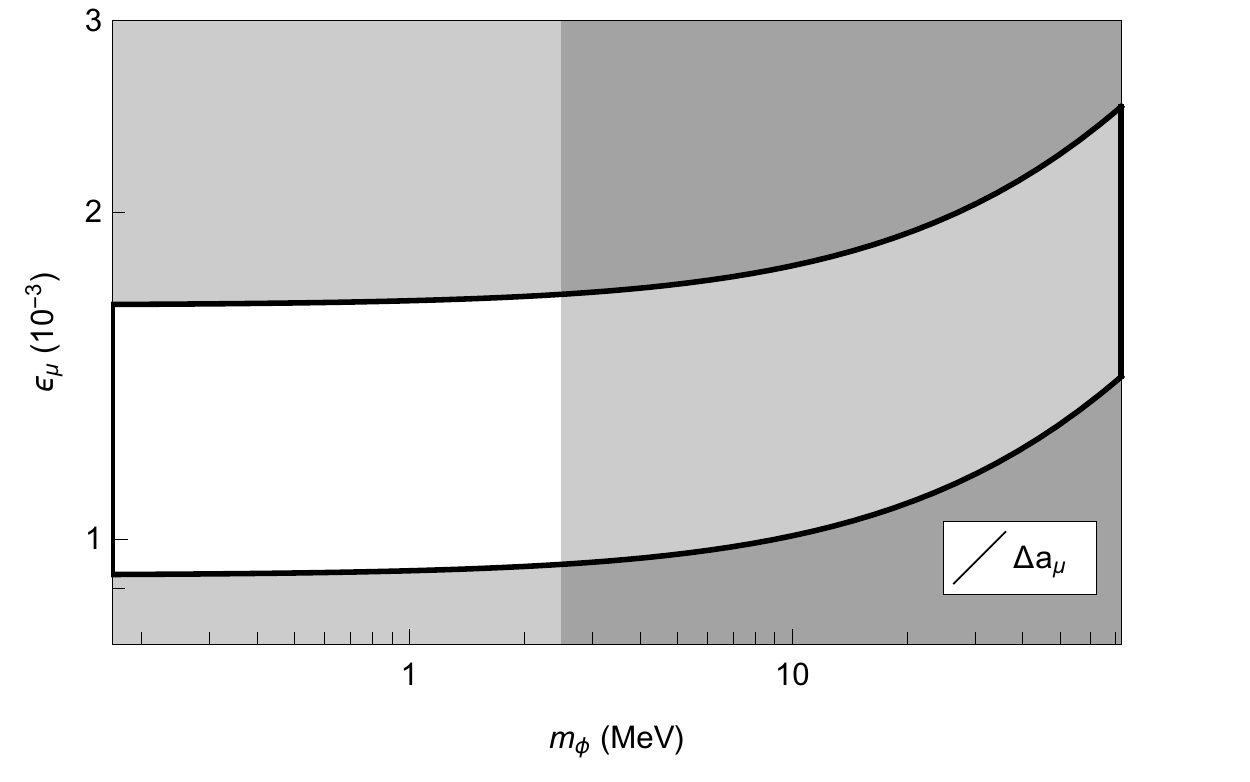}
\caption{\label{fig:gmu_exclusion+newH} Exclusion plot for $\epsilon_\mu$ (shaded region is excluded). See caption of Fig.~\ref{fig:gmu_exclusion} for more details.
}
\end{figure}

\begin{figure}[!ht]
\centering
\includegraphics[width=\columnwidth]{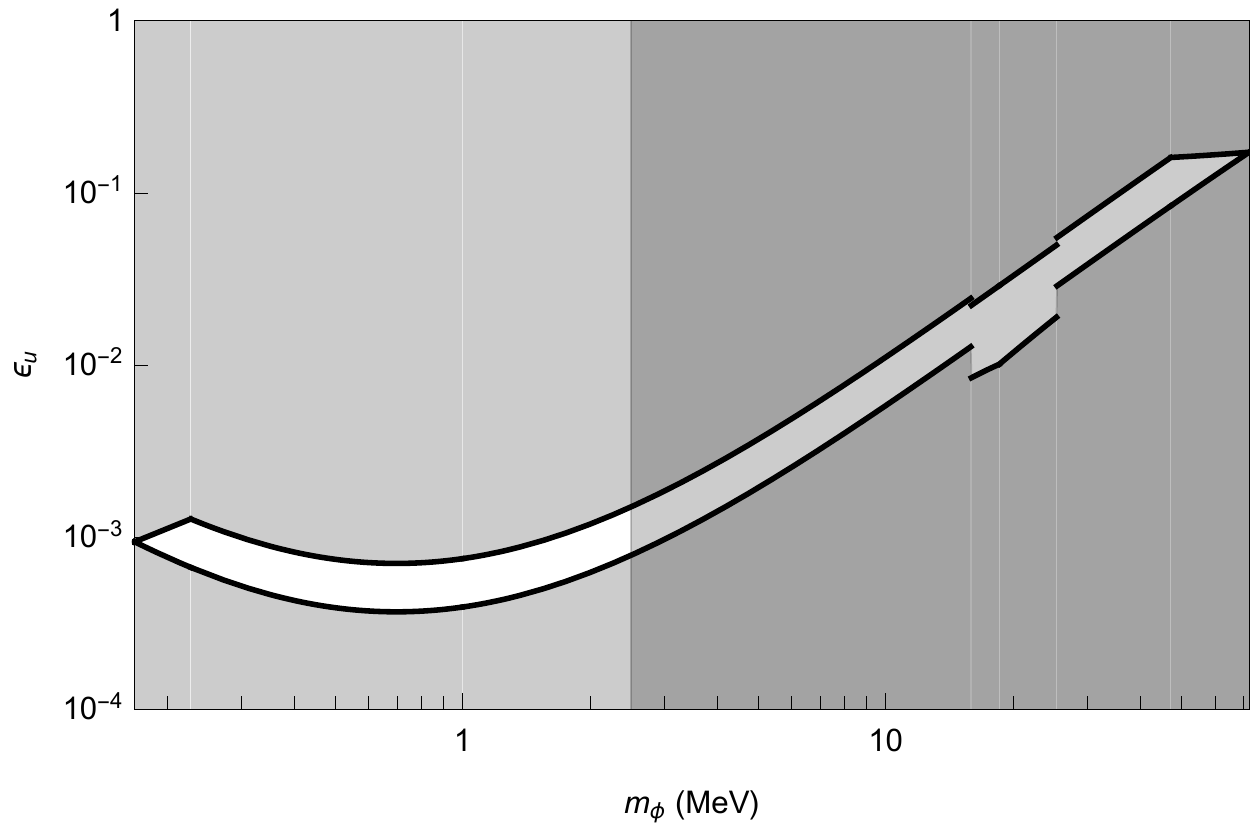}
\caption{\label{fig:gu_exclusion+newH} Exclusion plot for $\epsilon_u$ (shaded region is excluded). See caption of Fig.~\ref{fig:gu_exclusion} for more details.
}
\end{figure}

\begin{figure}[!ht]
\centering\includegraphics[width=\columnwidth]{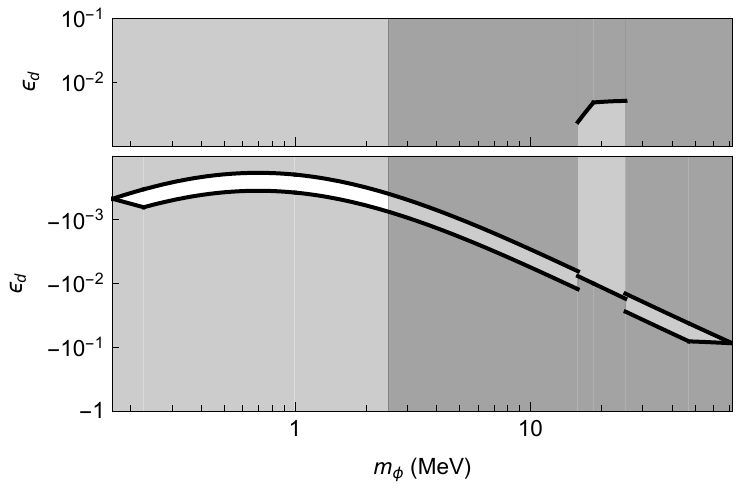}
\caption{\label{fig:gd_exclusion+newH} Exclusion plot for $\epsilon_d$ (shaded region is excluded). See caption of Fig.~\ref{fig:gd_exclusion} for more details.
}
\end{figure}

\begin{figure}[!ht]
\centering\includegraphics[width=\columnwidth]{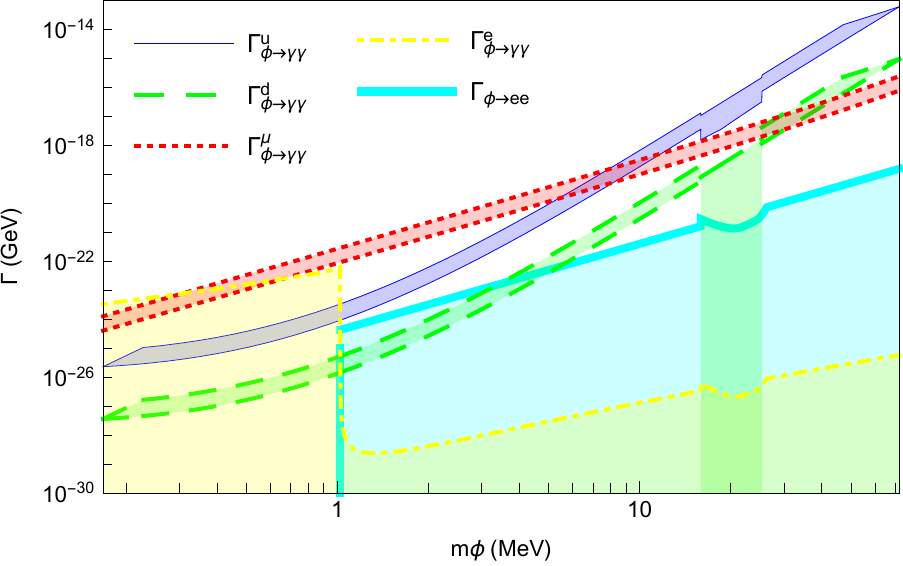}
\caption{\label{fig:phi_decay_rate+newH} The total $\phi$ decay width (shaded region is allowed to decay). See caption of Fig.~\ref{fig:phi_decay_rate} for more details.}
\end{figure}

\begin{figure}[!ht]
\centering\includegraphics[width=\columnwidth]{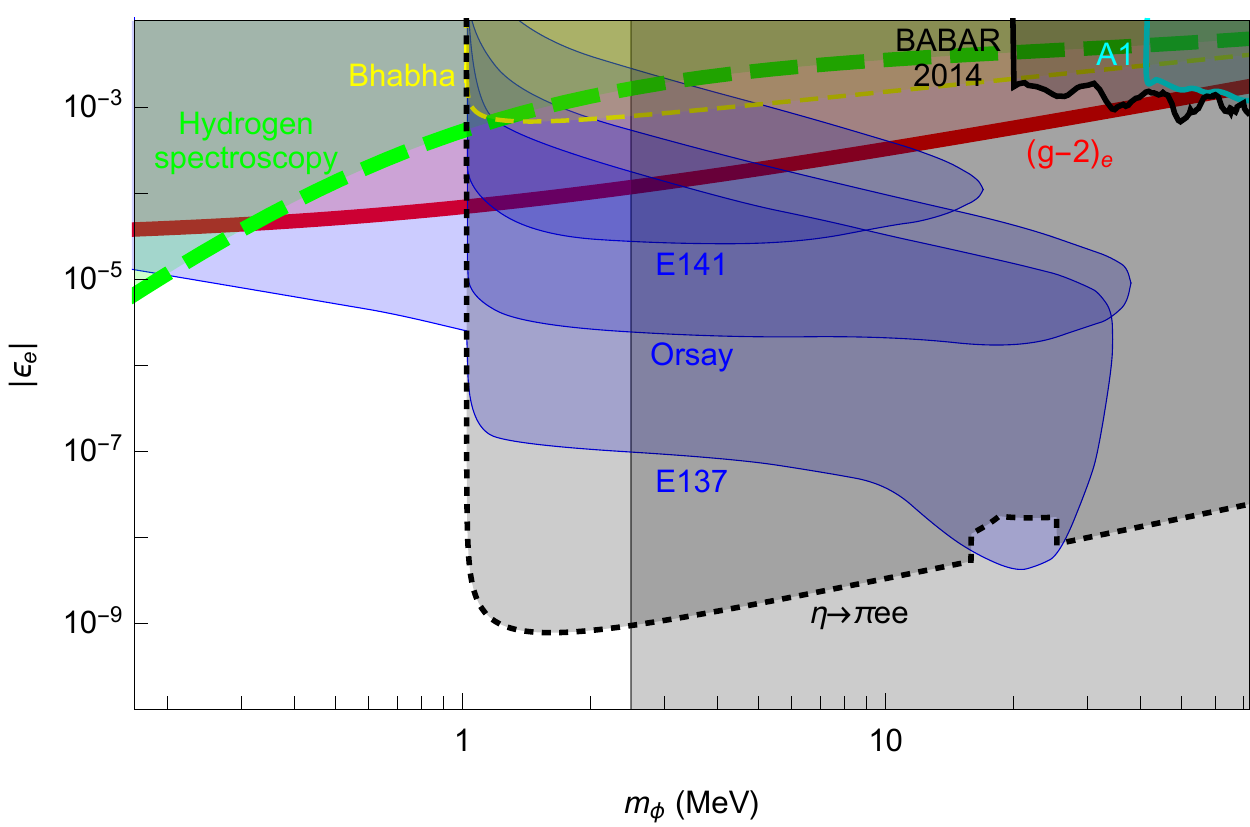}
\caption{\label{fig:ge_exclusion+newH} Exclusion plot for $\epsilon_e$ (shaded region is excluded). See caption of Fig.~\ref{fig:ge_exclusion} for more details.}
\end{figure}

\begin{figure}[!ht]
\centering\includegraphics[width=\columnwidth]{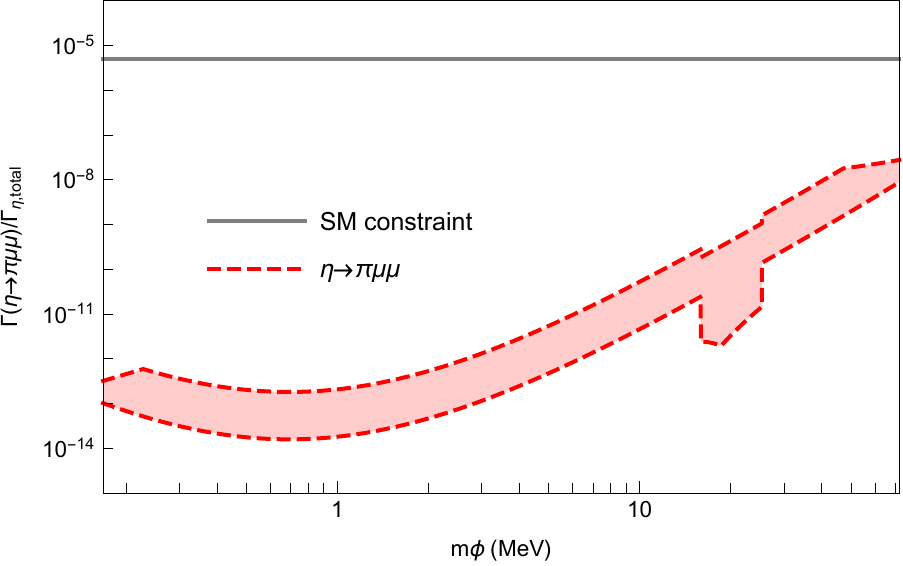}
\caption{\label{fig:eta_to_pi_mu_mu+newH} Decay rate of $\eta\to\pi^0\phi^*\to\pi^0\mu^+\mu^-$ (shaded region is allowed to decay). See caption of Fig.~\ref{fig:eta_to_pi_mu_mu} for more details.}
\end{figure}

\begin{figure}[!ht]
\centering\includegraphics[width=\columnwidth]{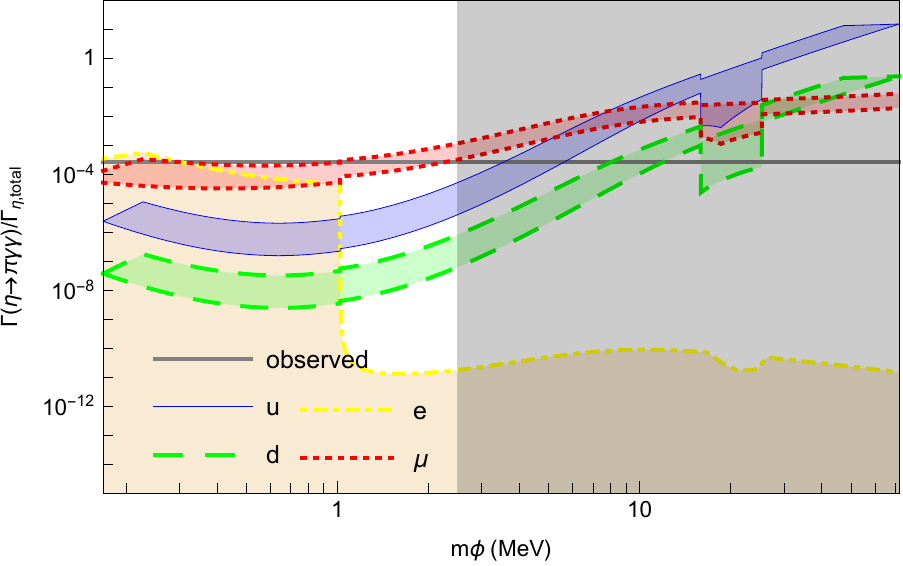}
\caption{\label{fig:eta_to_pi_gamma_gamma+newH} Decay rate of $\eta\to\pi^0\phi^*\to\pi^0\gamma\gamma$ (shaded region is allowed to decay). See caption of Fig.~\ref{fig:eta_to_pi_gamma_gamma} for more details.}
\end{figure}

To allow the new physics, the electron-proton scattering (CODATA value), and the new experiment to coexist, we assume that the actual proton radius lies within $3\sigma$ of both new and old experiments. Such a value is between 0.8568 to 0.8620\,fm. First, following our previous work, we obtain the constraint of $\epsilon_p$, $\epsilon_n$, $\epsilon_\mu$, $\epsilon_u$, and $\epsilon_d$, and the results are shown in Figs.~\ref{fig:gp_exclusion+newH}--\ref{fig:gd_exclusion+newH}, respectively. Second, we  repeat the analysis in Secs.~\ref{sec:eta to pi f f} and \ref{sec:eta to pi gamma gamma}, and obtain the $\phi$ decay rate in different channels in Fig.~\ref{fig:phi_decay_rate+newH}, $\epsilon_e$ exclusion plot in Fig.~\ref{fig:ge_exclusion+newH}, $\eta\to\pi^0\mu^+\mu^-$ decay rate in Fig.~\ref{fig:eta_to_pi_mu_mu+newH}, and $\eta\to\pi^0\gamma\gamma$ decay rate in different channels in Fig.~\ref{fig:eta_to_pi_gamma_gamma+newH}. Finally, we obtain the allowed $m_\phi$ is from 168 keV to 2.50 MeV and $\epsilon_u$, $\epsilon_p$, and $\epsilon_\mu$ are all around $10^{-3}$. Note that $\epsilon_n$ is completely ruled out with the new result and this means that the scalar coupling to the neutron is zero, \textit{i.e.} $\epsilon_u=-2\epsilon_d$.

At first glance, one might think that the effect of new physics is approximately halved, therefore the constraint should be less strict. This intuition is not correct. Comparing with the results in Sec.~\ref{sec:eta to pi gamma gamma}, we see that the upper bound of $m_\phi$ becomes smaller. The problem is that the effect of $\epsilon_n$ should be included carefully. To correctly analyze it, examine Eq.~(\ref{eq:eta to pi gamma gamma with NWA}). The $\eta\to\pi^0\gamma\gamma$ decay rate is proportional to $(\epsilon_u-\epsilon_d)^2$ (this factor comes from the $\eta\pi^0\phi$ vertex). In the previous scenario, $\epsilon_d$ can be positive (see Fig.~\ref{fig:gd_exclusion}), so $(\epsilon_u-\epsilon_d)^2$ can be smaller than $\epsilon_u^2$; in this section, $\epsilon_d$ stays negative in the allowed $m_\phi$ range (see Fig.~\ref{fig:gd_exclusion+newH}), so $(\epsilon_u-\epsilon_d)^2$ is bigger than $\epsilon_u^2$. The root cause of this strange behavior can be traced back to $\epsilon_n$ can be non-zero in the previous scenario, but strictly zero in this scenario.

%===============================================================================
%===============================================================================
\section{Conclusion}\label{sec:conclusion}
Our previous work limits the existence of the new scalar boson to the mass range of about 160 keV to 60 MeV. Here we reanalyze the beam dump experiments and find that the exclusion region is quite different than that of the previous work \cite{Liu:2016qwd}. With strong interaction input to $\eta$ decay from the NJL model, we present a tighter constraint on the new scalar boson $\phi$: The mass range is now from 160 keV to 3.5 MeV, $\epsilon_u$, $\epsilon_p$, and $\epsilon_\mu$ are all around $10^{-3}$, $\epsilon_n$ is from $-0.4$ to 0.2.

We also considered the scenario where the new physics coexists with the new regular hydrogen laser spectroscopy experiment and the old CODATA value. Most of the parameter space is similar, except $\epsilon_n$ is zero, meaning $\epsilon_u=-2\epsilon_d$.

One might expect that if we change the quark masses used in loop calculation in Sec. \ref{sec:phi decay rate} and \ref{sec:eta to pi gamma gamma}, the result might change drastically. However, this is not the case. Although the allowed decay channels in Fig. \ref{fig:eta_to_pi_gamma_gamma} and \ref{fig:eta_to_pi_gamma_gamma+newH} change accordingly when varying $u$ and $d$ quark masses, the resulting upper bound of $m_\phi$ is always around few MeV. After thorough investigation, we find the constraint of upper bound of $m_\phi$ is quite robust.

There are experiments aiming to explore $\eta$ decay with higher precision, such as recently approved the JLab Eta Factory (JEF) experiment \cite{Gan:2017kfr}, and proposed and the REDTOP project \cite{Gatto:2016rae}. There are several experiments that study the muonic puzzles: The MUSE experiment \cite{Gilman:2013eiv} plans to investigate the  proton radius puzzle by measuring $\mu^\pm$ and $e^\pm-p$ elastic scattering at low energies. The muon $g-2$ experiments at Fermilab \cite{Chapelain:2017syu} and J-PARC \cite{Iinuma:2016zfu} are of high interest for their own sake and for their bearing on the proton radius puzzle. The COMPASS collaboration is planning a radius measurement using their muon beam~\cite{Denisov:2018aa}.

The present work severely constrains the mass range of the possible new scalar boson $\phi$. Measurements aimed at investigating this particle for masses between 160 keV and 3.5 MeV could either discover the particle or completely rule it out. Our constraints are based on the assumption that only $\phi$ couples to SM particles through a simple Yukawa coupling. In the case of ruling out the scalar boson, this assumption becomes a constraint for model construction as an explanation for muonic puzzles. Indeed, there might be a subtle symmetry to forbid Yukawa terms, or there might be more complicated mechanisms for new physics to interact with the Standard Model.

\section*{Acknowledgments} 
We thank Liping Gan who suggests us to study $\eta$ decay to further constrain the new scalar boson. 
We also thank David McKeen and Yiming Zhong for valuable discussion.
The work of Y.-S. L. was supported by Science and Technology Commission of Shanghai Municipality (Grants No. 16DZ2260200) and National Natural Science Foundation of China (Grants No.11655002). The work of G.\,A.\,M. was supported by the U. S. Department of Energy Office of Science, Office of Nuclear Physics under Award Number DE-FG02-97ER-41014; and the work of I.\,C.\,C. was supported by the U.S. Department of Energy, Office of Science, Office of Nuclear Physics, contract no. DE-AC02-06CH11357.

%===============================================================================
%===============================================================================
\bibliographystyle{apsrev4-1}
\bibliography{bibliography}

\begin{thebibliography}{92}%
\makeatletter
\providecommand \@ifxundefined [1]{%
 \@ifx{#1\undefined}
}%
\providecommand \@ifnum [1]{%
 \ifnum #1\expandafter \@firstoftwo
 \else \expandafter \@secondoftwo
 \fi
}%
\providecommand \@ifx [1]{%
 \ifx #1\expandafter \@firstoftwo
 \else \expandafter \@secondoftwo
 \fi
}%
\providecommand \natexlab [1]{#1}%
\providecommand \enquote  [1]{``#1''}%
\providecommand \bibnamefont  [1]{#1}%
\providecommand \bibfnamefont [1]{#1}%
\providecommand \citenamefont [1]{#1}%
\providecommand \href@noop [0]{\@secondoftwo}%
\providecommand \href [0]{\begingroup \@sanitize@url \@href}%
\providecommand \@href[1]{\@@startlink{#1}\@@href}%
\providecommand \@@href[1]{\endgroup#1\@@endlink}%
\providecommand \@sanitize@url [0]{\catcode `\\12\catcode `\$12\catcode
  `\&12\catcode `\#12\catcode `\^12\catcode `\_12\catcode `\%12\relax}%
\providecommand \@@startlink[1]{}%
\providecommand \@@endlink[0]{}%
\providecommand \url  [0]{\begingroup\@sanitize@url \@url }%
\providecommand \@url [1]{\endgroup\@href {#1}{\urlprefix }}%
\providecommand \urlprefix  [0]{URL }%
\providecommand \Eprint [0]{\href }%
\providecommand \doibase [0]{http://dx.doi.org/}%
\providecommand \selectlanguage [0]{\@gobble}%
\providecommand \bibinfo  [0]{\@secondoftwo}%
\providecommand \bibfield  [0]{\@secondoftwo}%
\providecommand \translation [1]{[#1]}%
\providecommand \BibitemOpen [0]{}%
\providecommand \bibitemStop [0]{}%
\providecommand \bibitemNoStop [0]{.\EOS\space}%
\providecommand \EOS [0]{\spacefactor3000\relax}%
\providecommand \BibitemShut  [1]{\csname bibitem#1\endcsname}%
\let\auto@bib@innerbib\@empty
%</preamble>
\bibitem [{\citenamefont {Pohl}\ \emph {et~al.}(2010)\citenamefont {Pohl} \emph
  {et~al.}}]{Pohl:2010zza}%
  \BibitemOpen
  \bibfield  {author} {\bibinfo {author} {\bibfnamefont {R.}~\bibnamefont
  {Pohl}} \emph {et~al.},\ }\href {\doibase 10.1038/nature09250} {\bibfield
  {journal} {\bibinfo  {journal} {Nature}\ }\textbf {\bibinfo {volume} {466}},\
  \bibinfo {pages} {213} (\bibinfo {year} {2010})}\BibitemShut {NoStop}%
%%CITATION = NATUA,466,213;%%
\bibitem [{\citenamefont {Antognini}\ \emph {et~al.}(2013)\citenamefont
  {Antognini} \emph {et~al.}}]{Antognini:1900ns}%
  \BibitemOpen
  \bibfield  {author} {\bibinfo {author} {\bibfnamefont {A.}~\bibnamefont
  {Antognini}} \emph {et~al.},\ }\href {\doibase 10.1126/science.1230016}
  {\bibfield  {journal} {\bibinfo  {journal} {Science}\ }\textbf {\bibinfo
  {volume} {339}},\ \bibinfo {pages} {417} (\bibinfo {year}
  {2013})}\BibitemShut {NoStop}%
%%CITATION = SCIEA,339,417;%%
\bibitem [{\citenamefont {Mohr}\ \emph {et~al.}(2016)\citenamefont {Mohr},
  \citenamefont {Newell},\ and\ \citenamefont {Taylor}}]{Mohr:2015ccw}%
  \BibitemOpen
  \bibfield  {author} {\bibinfo {author} {\bibfnamefont {P.~J.}\ \bibnamefont
  {Mohr}}, \bibinfo {author} {\bibfnamefont {D.~B.}\ \bibnamefont {Newell}}, \
  and\ \bibinfo {author} {\bibfnamefont {B.~N.}\ \bibnamefont {Taylor}},\
  }\href {\doibase 10.1103/RevModPhys.88.035009} {\bibfield  {journal}
  {\bibinfo  {journal} {Rev. Mod. Phys.}\ }\textbf {\bibinfo {volume} {88}},\
  \bibinfo {pages} {035009} (\bibinfo {year} {2016})},\ \Eprint
  {http://arxiv.org/abs/1507.07956} {arXiv:1507.07956 [physics.atom-ph]}
  \BibitemShut {NoStop}%
%%CITATION = ARXIV:1507.07956;%%
\bibitem [{\citenamefont {Pohl}\ \emph {et~al.}(2013)\citenamefont {Pohl},
  \citenamefont {Gilman}, \citenamefont {Miller},\ and\ \citenamefont
  {Pachucki}}]{Pohl:2013yb}%
  \BibitemOpen
  \bibfield  {author} {\bibinfo {author} {\bibfnamefont {R.}~\bibnamefont
  {Pohl}}, \bibinfo {author} {\bibfnamefont {R.}~\bibnamefont {Gilman}},
  \bibinfo {author} {\bibfnamefont {G.~A.}\ \bibnamefont {Miller}}, \ and\
  \bibinfo {author} {\bibfnamefont {K.}~\bibnamefont {Pachucki}},\ }\href
  {\doibase 10.1146/annurev-nucl-102212-170627} {\bibfield  {journal} {\bibinfo
   {journal} {Ann. Rev. Nucl. Part. Sci.}\ }\textbf {\bibinfo {volume} {63}},\
  \bibinfo {pages} {175} (\bibinfo {year} {2013})},\ \Eprint
  {http://arxiv.org/abs/1301.0905} {arXiv:1301.0905 [physics.atom-ph]}
  \BibitemShut {NoStop}%
%%CITATION = ARXIV:1301.0905;%%
\bibitem [{\citenamefont {Carlson}(2015)}]{Carlson:2015jba}%
  \BibitemOpen
  \bibfield  {author} {\bibinfo {author} {\bibfnamefont {C.~E.}\ \bibnamefont
  {Carlson}},\ }\href {\doibase 10.1016/j.ppnp.2015.01.002} {\bibfield
  {journal} {\bibinfo  {journal} {Prog. Part. Nucl. Phys.}\ }\textbf {\bibinfo
  {volume} {82}},\ \bibinfo {pages} {59} (\bibinfo {year} {2015})},\ \Eprint
  {http://arxiv.org/abs/1502.05314} {arXiv:1502.05314 [hep-ph]} \BibitemShut
  {NoStop}%
%%CITATION = ARXIV:1502.05314;%%
\bibitem [{\citenamefont {Carlson}\ and\ \citenamefont
  {Rislow}(2012)}]{Carlson:2012pc}%
  \BibitemOpen
  \bibfield  {author} {\bibinfo {author} {\bibfnamefont {C.~E.}\ \bibnamefont
  {Carlson}}\ and\ \bibinfo {author} {\bibfnamefont {B.~C.}\ \bibnamefont
  {Rislow}},\ }\href {\doibase 10.1103/PhysRevD.86.035013} {\bibfield
  {journal} {\bibinfo  {journal} {Phys. Rev.}\ }\textbf {\bibinfo {volume}
  {D86}},\ \bibinfo {pages} {035013} (\bibinfo {year} {2012})},\ \Eprint
  {http://arxiv.org/abs/1206.3587} {arXiv:1206.3587 [hep-ph]} \BibitemShut
  {NoStop}%
%%CITATION = ARXIV:1206.3587;%%
\bibitem [{\citenamefont {Carlson}\ and\ \citenamefont
  {Rislow}(2014)}]{Carlson:2013mya}%
  \BibitemOpen
  \bibfield  {author} {\bibinfo {author} {\bibfnamefont {C.~E.}\ \bibnamefont
  {Carlson}}\ and\ \bibinfo {author} {\bibfnamefont {B.~C.}\ \bibnamefont
  {Rislow}},\ }\href {\doibase 10.1103/PhysRevD.89.035003} {\bibfield
  {journal} {\bibinfo  {journal} {Phys. Rev.}\ }\textbf {\bibinfo {volume}
  {D89}},\ \bibinfo {pages} {035003} (\bibinfo {year} {2014})},\ \Eprint
  {http://arxiv.org/abs/1310.2786} {arXiv:1310.2786 [hep-ph]} \BibitemShut
  {NoStop}%
%%CITATION = ARXIV:1310.2786;%%
\bibitem [{\citenamefont {Lindner}\ \emph {et~al.}(2018)\citenamefont
  {Lindner}, \citenamefont {Platscher},\ and\ \citenamefont
  {Queiroz}}]{Lindner:2016bgg}%
  \BibitemOpen
  \bibfield  {author} {\bibinfo {author} {\bibfnamefont {M.}~\bibnamefont
  {Lindner}}, \bibinfo {author} {\bibfnamefont {M.}~\bibnamefont {Platscher}},
  \ and\ \bibinfo {author} {\bibfnamefont {F.~S.}\ \bibnamefont {Queiroz}},\
  }\href {\doibase 10.1016/j.physrep.2017.12.001} {\bibfield  {journal}
  {\bibinfo  {journal} {Phys. Rept.}\ }\textbf {\bibinfo {volume} {731}},\
  \bibinfo {pages} {1} (\bibinfo {year} {2018})},\ \Eprint
  {http://arxiv.org/abs/1610.06587} {arXiv:1610.06587 [hep-ph]} \BibitemShut
  {NoStop}%
%%CITATION = ARXIV:1610.06587;%%
\bibitem [{\citenamefont {Blum}\ \emph {et~al.}(2013)\citenamefont {Blum},
  \citenamefont {Denig}, \citenamefont {Logashenko}, \citenamefont {de~Rafael},
  \citenamefont {Lee~Roberts}, \citenamefont {Teubner},\ and\ \citenamefont
  {Venanzoni}}]{Blum:2013xva}%
  \BibitemOpen
  \bibfield  {author} {\bibinfo {author} {\bibfnamefont {T.}~\bibnamefont
  {Blum}}, \bibinfo {author} {\bibfnamefont {A.}~\bibnamefont {Denig}},
  \bibinfo {author} {\bibfnamefont {I.}~\bibnamefont {Logashenko}}, \bibinfo
  {author} {\bibfnamefont {E.}~\bibnamefont {de~Rafael}}, \bibinfo {author}
  {\bibfnamefont {B.}~\bibnamefont {Lee~Roberts}}, \bibinfo {author}
  {\bibfnamefont {T.}~\bibnamefont {Teubner}}, \ and\ \bibinfo {author}
  {\bibfnamefont {G.}~\bibnamefont {Venanzoni}},\ }\href@noop {} {\  (\bibinfo
  {year} {2013})},\ \Eprint {http://arxiv.org/abs/1311.2198} {arXiv:1311.2198
  [hep-ph]} \BibitemShut {NoStop}%
%%CITATION = ARXIV:1311.2198;%%
\bibitem [{\citenamefont {Davier}\ \emph {et~al.}(2011)\citenamefont {Davier},
  \citenamefont {Hoecker}, \citenamefont {Malaescu},\ and\ \citenamefont
  {Zhang}}]{Davier:2010nc}%
  \BibitemOpen
  \bibfield  {author} {\bibinfo {author} {\bibfnamefont {M.}~\bibnamefont
  {Davier}}, \bibinfo {author} {\bibfnamefont {A.}~\bibnamefont {Hoecker}},
  \bibinfo {author} {\bibfnamefont {B.}~\bibnamefont {Malaescu}}, \ and\
  \bibinfo {author} {\bibfnamefont {Z.}~\bibnamefont {Zhang}},\ }\href
  {\doibase 10.1140/epjc/s10052-012-1874-8, 10.1140/epjc/s10052-010-1515-z}
  {\bibfield  {journal} {\bibinfo  {journal} {Eur. Phys. J.}\ }\textbf
  {\bibinfo {volume} {C71}},\ \bibinfo {pages} {1515} (\bibinfo {year}
  {2011})},\ \bibinfo {note} {[Erratum: Eur. Phys. J.C72,1874(2012)]},\ \Eprint
  {http://arxiv.org/abs/1010.4180} {arXiv:1010.4180 [hep-ph]} \BibitemShut
  {NoStop}%
%%CITATION = ARXIV:1010.4180;%%
\bibitem [{\citenamefont {Hagiwara}\ \emph {et~al.}(2011)\citenamefont
  {Hagiwara}, \citenamefont {Liao}, \citenamefont {Martin}, \citenamefont
  {Nomura},\ and\ \citenamefont {Teubner}}]{Hagiwara:2011af}%
  \BibitemOpen
  \bibfield  {author} {\bibinfo {author} {\bibfnamefont {K.}~\bibnamefont
  {Hagiwara}}, \bibinfo {author} {\bibfnamefont {R.}~\bibnamefont {Liao}},
  \bibinfo {author} {\bibfnamefont {A.~D.}\ \bibnamefont {Martin}}, \bibinfo
  {author} {\bibfnamefont {D.}~\bibnamefont {Nomura}}, \ and\ \bibinfo {author}
  {\bibfnamefont {T.}~\bibnamefont {Teubner}},\ }\href {\doibase
  10.1088/0954-3899/38/8/085003} {\bibfield  {journal} {\bibinfo  {journal} {J.
  Phys.}\ }\textbf {\bibinfo {volume} {G38}},\ \bibinfo {pages} {085003}
  (\bibinfo {year} {2011})},\ \Eprint {http://arxiv.org/abs/1105.3149}
  {arXiv:1105.3149 [hep-ph]} \BibitemShut {NoStop}%
%%CITATION = ARXIV:1105.3149;%%
\bibitem [{\citenamefont {Liu}(2017)}]{Liu:2017bzj}%
  \BibitemOpen
  \bibfield  {author} {\bibinfo {author} {\bibfnamefont {Y.-S.}\ \bibnamefont
  {Liu}},\ }\emph {\bibinfo {title} {{Possible Beyond the Standard Model
  Physics Motivated by Muonic Puzzles}}},\ \href
  {https://inspirehep.net/record/1614639/files/arXiv:1708.01655.pdf} {Ph.D.
  thesis},\ \bibinfo  {school} {Washington U., Seattle} (\bibinfo {year}
  {2017}),\ \Eprint {http://arxiv.org/abs/1708.01655} {arXiv:1708.01655
  [hep-ph]} \BibitemShut {NoStop}%
%%CITATION = ARXIV:1708.01655;%%
\bibitem [{\citenamefont {Tucker-Smith}\ and\ \citenamefont
  {Yavin}(2011)}]{TuckerSmith:2010ra}%
  \BibitemOpen
  \bibfield  {author} {\bibinfo {author} {\bibfnamefont {D.}~\bibnamefont
  {Tucker-Smith}}\ and\ \bibinfo {author} {\bibfnamefont {I.}~\bibnamefont
  {Yavin}},\ }\href {\doibase 10.1103/PhysRevD.83.101702} {\bibfield  {journal}
  {\bibinfo  {journal} {Phys. Rev.}\ }\textbf {\bibinfo {volume} {D83}},\
  \bibinfo {pages} {101702} (\bibinfo {year} {2011})},\ \Eprint
  {http://arxiv.org/abs/1011.4922} {arXiv:1011.4922 [hep-ph]} \BibitemShut
  {NoStop}%
%%CITATION = ARXIV:1011.4922;%%
\bibitem [{\citenamefont {Liu}\ and\ \citenamefont
  {Miller}(2015)}]{Liu:2015sba}%
  \BibitemOpen
  \bibfield  {author} {\bibinfo {author} {\bibfnamefont {Y.-S.}\ \bibnamefont
  {Liu}}\ and\ \bibinfo {author} {\bibfnamefont {G.~A.}\ \bibnamefont
  {Miller}},\ }\href {\doibase 10.1103/PhysRevC.92.035209} {\bibfield
  {journal} {\bibinfo  {journal} {Phys. Rev.}\ }\textbf {\bibinfo {volume}
  {C92}},\ \bibinfo {pages} {035209} (\bibinfo {year} {2015})},\ \Eprint
  {http://arxiv.org/abs/1507.04399} {arXiv:1507.04399 [nucl-th]} \BibitemShut
  {NoStop}%
%%CITATION = ARXIV:1507.04399;%%
\bibitem [{\citenamefont {Kinoshita}\ and\ \citenamefont
  {Marciano}(1990)}]{Kinoshita:1990aj}%
  \BibitemOpen
  \bibfield  {author} {\bibinfo {author} {\bibfnamefont {T.}~\bibnamefont
  {Kinoshita}}\ and\ \bibinfo {author} {\bibfnamefont {W.~J.}\ \bibnamefont
  {Marciano}},\ }\href {\doibase 10.1142/9789814503273_0010} {\bibfield
  {journal} {\bibinfo  {journal} {Adv. Ser. Direct. High Energy Phys.}\
  }\textbf {\bibinfo {volume} {7}},\ \bibinfo {pages} {419} (\bibinfo {year}
  {1990})}\BibitemShut {NoStop}%
%%CITATION = 00319,7,419;%%
\bibitem [{\citenamefont {Batell}\ \emph {et~al.}(2011)\citenamefont {Batell},
  \citenamefont {McKeen},\ and\ \citenamefont {Pospelov}}]{Batell:2011qq}%
  \BibitemOpen
  \bibfield  {author} {\bibinfo {author} {\bibfnamefont {B.}~\bibnamefont
  {Batell}}, \bibinfo {author} {\bibfnamefont {D.}~\bibnamefont {McKeen}}, \
  and\ \bibinfo {author} {\bibfnamefont {M.}~\bibnamefont {Pospelov}},\ }\href
  {\doibase 10.1103/PhysRevLett.107.011803} {\bibfield  {journal} {\bibinfo
  {journal} {Phys. Rev. Lett.}\ }\textbf {\bibinfo {volume} {107}},\ \bibinfo
  {pages} {011803} (\bibinfo {year} {2011})},\ \Eprint
  {http://arxiv.org/abs/1103.0721} {arXiv:1103.0721 [hep-ph]} \BibitemShut
  {NoStop}%
%%CITATION = ARXIV:1103.0721;%%
\bibitem [{\citenamefont {Schmidt-Hoberg}\ \emph {et~al.}(2013)\citenamefont
  {Schmidt-Hoberg}, \citenamefont {Staub},\ and\ \citenamefont
  {Winkler}}]{Schmidt-Hoberg:2013hba}%
  \BibitemOpen
  \bibfield  {author} {\bibinfo {author} {\bibfnamefont {K.}~\bibnamefont
  {Schmidt-Hoberg}}, \bibinfo {author} {\bibfnamefont {F.}~\bibnamefont
  {Staub}}, \ and\ \bibinfo {author} {\bibfnamefont {M.~W.}\ \bibnamefont
  {Winkler}},\ }\href {\doibase 10.1016/j.physletb.2013.11.015} {\bibfield
  {journal} {\bibinfo  {journal} {Phys. Lett.}\ }\textbf {\bibinfo {volume}
  {B727}},\ \bibinfo {pages} {506} (\bibinfo {year} {2013})},\ \Eprint
  {http://arxiv.org/abs/1310.6752} {arXiv:1310.6752 [hep-ph]} \BibitemShut
  {NoStop}%
%%CITATION = ARXIV:1310.6752;%%
\bibitem [{\citenamefont {Essig}\ \emph {et~al.}(2013)\citenamefont {Essig}
  \emph {et~al.}}]{Essig:2013lka}%
  \BibitemOpen
  \bibfield  {author} {\bibinfo {author} {\bibfnamefont {R.}~\bibnamefont
  {Essig}} \emph {et~al.},\ }in\ \href
  {https://inspirehep.net/record/1263039/files/arXiv:1311.0029.pdf} {\emph
  {\bibinfo {booktitle} {{Proceedings, 2013 Community Summer Study on the
  Future of U.S. Particle Physics: Snowmass on the Mississippi (CSS2013):
  Minneapolis, MN, USA, July 29-August 6, 2013}}}}\ (\bibinfo {year} {2013})\
  \Eprint {http://arxiv.org/abs/1311.0029} {arXiv:1311.0029 [hep-ph]}
  \BibitemShut {NoStop}%
%%CITATION = ARXIV:1311.0029;%%
\bibitem [{\citenamefont {Knapen}\ and\ \citenamefont
  {Robinson}(2015)}]{Knapen:2015hia}%
  \BibitemOpen
  \bibfield  {author} {\bibinfo {author} {\bibfnamefont {S.}~\bibnamefont
  {Knapen}}\ and\ \bibinfo {author} {\bibfnamefont {D.~J.}\ \bibnamefont
  {Robinson}},\ }\href {\doibase 10.1103/PhysRevLett.115.161803} {\bibfield
  {journal} {\bibinfo  {journal} {Phys. Rev. Lett.}\ }\textbf {\bibinfo
  {volume} {115}},\ \bibinfo {pages} {161803} (\bibinfo {year} {2015})},\
  \Eprint {http://arxiv.org/abs/1507.00009} {arXiv:1507.00009 [hep-ph]}
  \BibitemShut {NoStop}%
%%CITATION = ARXIV:1507.00009;%%
\bibitem [{\citenamefont {Chen}\ \emph {et~al.}(2016)\citenamefont {Chen},
  \citenamefont {Davoudiasl}, \citenamefont {Marciano},\ and\ \citenamefont
  {Zhang}}]{Chen:2015vqy}%
  \BibitemOpen
  \bibfield  {author} {\bibinfo {author} {\bibfnamefont {C.-Y.}\ \bibnamefont
  {Chen}}, \bibinfo {author} {\bibfnamefont {H.}~\bibnamefont {Davoudiasl}},
  \bibinfo {author} {\bibfnamefont {W.~J.}\ \bibnamefont {Marciano}}, \ and\
  \bibinfo {author} {\bibfnamefont {C.}~\bibnamefont {Zhang}},\ }\href
  {\doibase 10.1103/PhysRevD.93.035006} {\bibfield  {journal} {\bibinfo
  {journal} {Phys. Rev.}\ }\textbf {\bibinfo {volume} {D93}},\ \bibinfo {pages}
  {035006} (\bibinfo {year} {2016})},\ \Eprint
  {http://arxiv.org/abs/1511.04715} {arXiv:1511.04715 [hep-ph]} \BibitemShut
  {NoStop}%
%%CITATION = ARXIV:1511.04715;%%
\bibitem [{\citenamefont {Batell}\ \emph {et~al.}(2017)\citenamefont {Batell},
  \citenamefont {Lange}, \citenamefont {McKeen}, \citenamefont {Pospelov},\
  and\ \citenamefont {Ritz}}]{Batell:2016ove}%
  \BibitemOpen
  \bibfield  {author} {\bibinfo {author} {\bibfnamefont {B.}~\bibnamefont
  {Batell}}, \bibinfo {author} {\bibfnamefont {N.}~\bibnamefont {Lange}},
  \bibinfo {author} {\bibfnamefont {D.}~\bibnamefont {McKeen}}, \bibinfo
  {author} {\bibfnamefont {M.}~\bibnamefont {Pospelov}}, \ and\ \bibinfo
  {author} {\bibfnamefont {A.}~\bibnamefont {Ritz}},\ }\href {\doibase
  10.1103/PhysRevD.95.075003} {\bibfield  {journal} {\bibinfo  {journal} {Phys.
  Rev.}\ }\textbf {\bibinfo {volume} {D95}},\ \bibinfo {pages} {075003}
  (\bibinfo {year} {2017})},\ \Eprint {http://arxiv.org/abs/1606.04943}
  {arXiv:1606.04943 [hep-ph]} \BibitemShut {NoStop}%
%%CITATION = ARXIV:1606.04943;%%
\bibitem [{\citenamefont {Batell}\ \emph {et~al.}(2018)\citenamefont {Batell},
  \citenamefont {Freitas}, \citenamefont {Ismail},\ and\ \citenamefont
  {Mckeen}}]{Batell:2017kty}%
  \BibitemOpen
  \bibfield  {author} {\bibinfo {author} {\bibfnamefont {B.}~\bibnamefont
  {Batell}}, \bibinfo {author} {\bibfnamefont {A.}~\bibnamefont {Freitas}},
  \bibinfo {author} {\bibfnamefont {A.}~\bibnamefont {Ismail}}, \ and\ \bibinfo
  {author} {\bibfnamefont {D.}~\bibnamefont {Mckeen}},\ }\href {\doibase
  10.1103/PhysRevD.98.055026} {\bibfield  {journal} {\bibinfo  {journal} {Phys.
  Rev.}\ }\textbf {\bibinfo {volume} {D98}},\ \bibinfo {pages} {055026}
  (\bibinfo {year} {2018})},\ \Eprint {http://arxiv.org/abs/1712.10022}
  {arXiv:1712.10022 [hep-ph]} \BibitemShut {NoStop}%
%%CITATION = ARXIV:1712.10022;%%
\bibitem [{\citenamefont {Izaguirre}\ \emph {et~al.}(2015)\citenamefont
  {Izaguirre}, \citenamefont {Krnjaic},\ and\ \citenamefont
  {Pospelov}}]{Izaguirre:2014cza}%
  \BibitemOpen
  \bibfield  {author} {\bibinfo {author} {\bibfnamefont {E.}~\bibnamefont
  {Izaguirre}}, \bibinfo {author} {\bibfnamefont {G.}~\bibnamefont {Krnjaic}},
  \ and\ \bibinfo {author} {\bibfnamefont {M.}~\bibnamefont {Pospelov}},\
  }\href {\doibase 10.1016/j.physletb.2014.11.037} {\bibfield  {journal}
  {\bibinfo  {journal} {Phys. Lett.}\ }\textbf {\bibinfo {volume} {B740}},\
  \bibinfo {pages} {61} (\bibinfo {year} {2015})},\ \Eprint
  {http://arxiv.org/abs/1405.4864} {arXiv:1405.4864 [hep-ph]} \BibitemShut
  {NoStop}%
%%CITATION = ARXIV:1405.4864;%%
\bibitem [{\citenamefont {Chen}\ \emph {et~al.}(2017)\citenamefont {Chen},
  \citenamefont {Pospelov},\ and\ \citenamefont {Zhong}}]{Chen:2017awl}%
  \BibitemOpen
  \bibfield  {author} {\bibinfo {author} {\bibfnamefont {C.-Y.}\ \bibnamefont
  {Chen}}, \bibinfo {author} {\bibfnamefont {M.}~\bibnamefont {Pospelov}}, \
  and\ \bibinfo {author} {\bibfnamefont {Y.-M.}\ \bibnamefont {Zhong}},\ }\href
  {\doibase 10.1103/PhysRevD.95.115005} {\bibfield  {journal} {\bibinfo
  {journal} {Phys. Rev.}\ }\textbf {\bibinfo {volume} {D95}},\ \bibinfo {pages}
  {115005} (\bibinfo {year} {2017})},\ \Eprint
  {http://arxiv.org/abs/1701.07437} {arXiv:1701.07437 [hep-ph]} \BibitemShut
  {NoStop}%
%%CITATION = ARXIV:1701.07437;%%
\bibitem [{\citenamefont {Chen}\ \emph {et~al.}(2018)\citenamefont {Chen},
  \citenamefont {Kozaczuk},\ and\ \citenamefont {Zhong}}]{Chen:2018vkr}%
  \BibitemOpen
  \bibfield  {author} {\bibinfo {author} {\bibfnamefont {C.-Y.}\ \bibnamefont
  {Chen}}, \bibinfo {author} {\bibfnamefont {J.}~\bibnamefont {Kozaczuk}}, \
  and\ \bibinfo {author} {\bibfnamefont {Y.-M.}\ \bibnamefont {Zhong}},\
  }\href@noop {} {\  (\bibinfo {year} {2018})},\ \Eprint
  {http://arxiv.org/abs/1807.03790} {arXiv:1807.03790 [hep-ph]} \BibitemShut
  {NoStop}%
%%CITATION = ARXIV:1807.03790;%%
\bibitem [{\citenamefont {Kahn}\ \emph {et~al.}(2018)\citenamefont {Kahn},
  \citenamefont {Krnjaic}, \citenamefont {Tran},\ and\ \citenamefont
  {Whitbeck}}]{Kahn:2018cqs}%
  \BibitemOpen
  \bibfield  {author} {\bibinfo {author} {\bibfnamefont {Y.}~\bibnamefont
  {Kahn}}, \bibinfo {author} {\bibfnamefont {G.}~\bibnamefont {Krnjaic}},
  \bibinfo {author} {\bibfnamefont {N.}~\bibnamefont {Tran}}, \ and\ \bibinfo
  {author} {\bibfnamefont {A.}~\bibnamefont {Whitbeck}},\ }\href@noop {} {\
  (\bibinfo {year} {2018})},\ \Eprint {http://arxiv.org/abs/1804.03144}
  {arXiv:1804.03144 [hep-ph]} \BibitemShut {NoStop}%
%%CITATION = ARXIV:1804.03144;%%
\bibitem [{\citenamefont {Berlin}\ \emph {et~al.}(2018)\citenamefont {Berlin},
  \citenamefont {Gori}, \citenamefont {Schuster},\ and\ \citenamefont
  {Toro}}]{Berlin:2018pwi}%
  \BibitemOpen
  \bibfield  {author} {\bibinfo {author} {\bibfnamefont {A.}~\bibnamefont
  {Berlin}}, \bibinfo {author} {\bibfnamefont {S.}~\bibnamefont {Gori}},
  \bibinfo {author} {\bibfnamefont {P.}~\bibnamefont {Schuster}}, \ and\
  \bibinfo {author} {\bibfnamefont {N.}~\bibnamefont {Toro}},\ }\href {\doibase
  10.1103/PhysRevD.98.035011} {\bibfield  {journal} {\bibinfo  {journal} {Phys.
  Rev.}\ }\textbf {\bibinfo {volume} {D98}},\ \bibinfo {pages} {035011}
  (\bibinfo {year} {2018})},\ \Eprint {http://arxiv.org/abs/1804.00661}
  {arXiv:1804.00661 [hep-ph]} \BibitemShut {NoStop}%
%%CITATION = ARXIV:1804.00661;%%
\bibitem [{\citenamefont {Pospelov}\ and\ \citenamefont
  {Tsai}(2018)}]{Pospelov:2017kep}%
  \BibitemOpen
  \bibfield  {author} {\bibinfo {author} {\bibfnamefont {M.}~\bibnamefont
  {Pospelov}}\ and\ \bibinfo {author} {\bibfnamefont {Y.-D.}\ \bibnamefont
  {Tsai}},\ }\href {\doibase 10.1016/j.physletb.2018.08.053} {\bibfield
  {journal} {\bibinfo  {journal} {Phys. Lett.}\ }\textbf {\bibinfo {volume}
  {B785}},\ \bibinfo {pages} {288} (\bibinfo {year} {2018})},\ \Eprint
  {http://arxiv.org/abs/1706.00424} {arXiv:1706.00424 [hep-ph]} \BibitemShut
  {NoStop}%
%%CITATION = ARXIV:1706.00424;%%
\bibitem [{\citenamefont {Liu}\ \emph {et~al.}(2017)\citenamefont {Liu},
  \citenamefont {McKeen},\ and\ \citenamefont {Miller}}]{Liu:2016mqv}%
  \BibitemOpen
  \bibfield  {author} {\bibinfo {author} {\bibfnamefont {Y.-S.}\ \bibnamefont
  {Liu}}, \bibinfo {author} {\bibfnamefont {D.}~\bibnamefont {McKeen}}, \ and\
  \bibinfo {author} {\bibfnamefont {G.~A.}\ \bibnamefont {Miller}},\ }\href
  {\doibase 10.1103/PhysRevD.95.036010} {\bibfield  {journal} {\bibinfo
  {journal} {Phys. Rev.}\ }\textbf {\bibinfo {volume} {D95}},\ \bibinfo {pages}
  {036010} (\bibinfo {year} {2017})},\ \Eprint
  {http://arxiv.org/abs/1609.06781} {arXiv:1609.06781 [hep-ph]} \BibitemShut
  {NoStop}%
%%CITATION = ARXIV:1609.06781;%%
\bibitem [{\citenamefont {Liu}\ and\ \citenamefont
  {Miller}(2017)}]{Liu:2017htz}%
  \BibitemOpen
  \bibfield  {author} {\bibinfo {author} {\bibfnamefont {Y.-S.}\ \bibnamefont
  {Liu}}\ and\ \bibinfo {author} {\bibfnamefont {G.~A.}\ \bibnamefont
  {Miller}},\ }\href {\doibase 10.1103/PhysRevD.96.016004} {\bibfield
  {journal} {\bibinfo  {journal} {Phys. Rev.}\ }\textbf {\bibinfo {volume}
  {D96}},\ \bibinfo {pages} {016004} (\bibinfo {year} {2017})},\ \Eprint
  {http://arxiv.org/abs/1705.01633} {arXiv:1705.01633 [hep-ph]} \BibitemShut
  {NoStop}%
%%CITATION = ARXIV:1705.01633;%%
\bibitem [{\citenamefont {Bjorken}\ \emph {et~al.}(2009)\citenamefont
  {Bjorken}, \citenamefont {Essig}, \citenamefont {Schuster},\ and\
  \citenamefont {Toro}}]{Bjorken:2009mm}%
  \BibitemOpen
  \bibfield  {author} {\bibinfo {author} {\bibfnamefont {J.~D.}\ \bibnamefont
  {Bjorken}}, \bibinfo {author} {\bibfnamefont {R.}~\bibnamefont {Essig}},
  \bibinfo {author} {\bibfnamefont {P.}~\bibnamefont {Schuster}}, \ and\
  \bibinfo {author} {\bibfnamefont {N.}~\bibnamefont {Toro}},\ }\href {\doibase
  10.1103/PhysRevD.80.075018} {\bibfield  {journal} {\bibinfo  {journal} {Phys.
  Rev.}\ }\textbf {\bibinfo {volume} {D80}},\ \bibinfo {pages} {075018}
  (\bibinfo {year} {2009})},\ \Eprint {http://arxiv.org/abs/0906.0580}
  {arXiv:0906.0580 [hep-ph]} \BibitemShut {NoStop}%
%%CITATION = ARXIV:0906.0580;%%
\bibitem [{\citenamefont {Andreas}\ \emph {et~al.}(2012)\citenamefont
  {Andreas}, \citenamefont {Niebuhr},\ and\ \citenamefont
  {Ringwald}}]{Andreas:2012mt}%
  \BibitemOpen
  \bibfield  {author} {\bibinfo {author} {\bibfnamefont {S.}~\bibnamefont
  {Andreas}}, \bibinfo {author} {\bibfnamefont {C.}~\bibnamefont {Niebuhr}}, \
  and\ \bibinfo {author} {\bibfnamefont {A.}~\bibnamefont {Ringwald}},\ }\href
  {\doibase 10.1103/PhysRevD.86.095019} {\bibfield  {journal} {\bibinfo
  {journal} {Phys. Rev.}\ }\textbf {\bibinfo {volume} {D86}},\ \bibinfo {pages}
  {095019} (\bibinfo {year} {2012})},\ \Eprint {http://arxiv.org/abs/1209.6083}
  {arXiv:1209.6083 [hep-ph]} \BibitemShut {NoStop}%
%%CITATION = ARXIV:1209.6083;%%
\bibitem [{\citenamefont {Nambu}\ and\ \citenamefont
  {Jona-Lasinio}(1961{\natexlab{a}})}]{Nambu:1961fr}%
  \BibitemOpen
  \bibfield  {author} {\bibinfo {author} {\bibfnamefont {Y.}~\bibnamefont
  {Nambu}}\ and\ \bibinfo {author} {\bibfnamefont {G.}~\bibnamefont
  {Jona-Lasinio}},\ }\href {\doibase 10.1103/PhysRev.124.246} {\bibfield
  {journal} {\bibinfo  {journal} {Phys. Rev.}\ }\textbf {\bibinfo {volume}
  {124}},\ \bibinfo {pages} {246} (\bibinfo {year}
  {1961}{\natexlab{a}})}\BibitemShut {NoStop}%
%%CITATION = PHRVA,124,246;%%
\bibitem [{\citenamefont {Nambu}\ and\ \citenamefont
  {Jona-Lasinio}(1961{\natexlab{b}})}]{Nambu:1961tp}%
  \BibitemOpen
  \bibfield  {author} {\bibinfo {author} {\bibfnamefont {Y.}~\bibnamefont
  {Nambu}}\ and\ \bibinfo {author} {\bibfnamefont {G.}~\bibnamefont
  {Jona-Lasinio}},\ }\href {\doibase 10.1103/PhysRev.122.345} {\bibfield
  {journal} {\bibinfo  {journal} {Phys. Rev.}\ }\textbf {\bibinfo {volume}
  {122}},\ \bibinfo {pages} {345} (\bibinfo {year}
  {1961}{\natexlab{b}})}\BibitemShut {NoStop}%
%%CITATION = PHRVA,122,345;%%
\bibitem [{\citenamefont {Vogl}\ and\ \citenamefont
  {Weise}(1991)}]{Vogl:1991qt}%
  \BibitemOpen
  \bibfield  {author} {\bibinfo {author} {\bibfnamefont {U.}~\bibnamefont
  {Vogl}}\ and\ \bibinfo {author} {\bibfnamefont {W.}~\bibnamefont {Weise}},\
  }\href {\doibase 10.1016/0146-6410(91)90005-9} {\bibfield  {journal}
  {\bibinfo  {journal} {Prog. Part. Nucl. Phys.}\ }\textbf {\bibinfo {volume}
  {27}},\ \bibinfo {pages} {195} (\bibinfo {year} {1991})}\BibitemShut
  {NoStop}%
%%CITATION = PPNPD,27,195;%%
\bibitem [{\citenamefont {Klevansky}(1992)}]{Klevansky:1992qe}%
  \BibitemOpen
  \bibfield  {author} {\bibinfo {author} {\bibfnamefont {S.}~\bibnamefont
  {Klevansky}},\ }\href {\doibase 10.1103/RevModPhys.64.649} {\bibfield
  {journal} {\bibinfo  {journal} {Rev. Mod. Phys.}\ }\textbf {\bibinfo {volume}
  {64}},\ \bibinfo {pages} {649} (\bibinfo {year} {1992})}\BibitemShut
  {NoStop}%
%%CITATION = RMPHA,64,649;%%
\bibitem [{\citenamefont {Beyer}\ \emph {et~al.}(2017)\citenamefont {Beyer},
  \citenamefont {Maisenbacher}, \citenamefont {Matveev}, \citenamefont {Pohl},
  \citenamefont {Khabarova}, \citenamefont {Grinin}, \citenamefont {Lamour},
  \citenamefont {Yost}, \citenamefont {H{\"a}nsch}, \citenamefont
  {Kolachevsky},\ and\ \citenamefont {Udem}}]{Beyer79}%
  \BibitemOpen
  \bibfield  {author} {\bibinfo {author} {\bibfnamefont {A.}~\bibnamefont
  {Beyer}}, \bibinfo {author} {\bibfnamefont {L.}~\bibnamefont {Maisenbacher}},
  \bibinfo {author} {\bibfnamefont {A.}~\bibnamefont {Matveev}}, \bibinfo
  {author} {\bibfnamefont {R.}~\bibnamefont {Pohl}}, \bibinfo {author}
  {\bibfnamefont {K.}~\bibnamefont {Khabarova}}, \bibinfo {author}
  {\bibfnamefont {A.}~\bibnamefont {Grinin}}, \bibinfo {author} {\bibfnamefont
  {T.}~\bibnamefont {Lamour}}, \bibinfo {author} {\bibfnamefont {D.~C.}\
  \bibnamefont {Yost}}, \bibinfo {author} {\bibfnamefont {T.~W.}\ \bibnamefont
  {H{\"a}nsch}}, \bibinfo {author} {\bibfnamefont {N.}~\bibnamefont
  {Kolachevsky}}, \ and\ \bibinfo {author} {\bibfnamefont {T.}~\bibnamefont
  {Udem}},\ }\href {\doibase 10.1126/science.aah6677} {\bibfield  {journal}
  {\bibinfo  {journal} {Science}\ }\textbf {\bibinfo {volume} {358}},\ \bibinfo
  {pages} {79} (\bibinfo {year} {2017})},\ \Eprint
  {http://arxiv.org/abs/http://science.sciencemag.org/content/358/6359/79.full.pdf}
  {http://science.sciencemag.org/content/358/6359/79.full.pdf} \BibitemShut
  {NoStop}%
\bibitem [{Ant()}]{Antognini}%
  \BibitemOpen
  \href@noop {} {\bibinfo  {journal} {Reported at 2017 Fall Meeting of the APS
  Division of Nuclear Physics by Aldo Antognini.}\ }\BibitemShut {NoStop}%
\bibitem [{\citenamefont {Sick}(2015)}]{Sick:2015spa}%
  \BibitemOpen
\bibfield  {journal} {  }\bibfield  {author} {\bibinfo {author} {\bibfnamefont
  {I.}~\bibnamefont {Sick}},\ }\href@noop {} {\  (\bibinfo {year} {2015})},\
  \Eprint {http://arxiv.org/abs/1505.06924} {arXiv:1505.06924 [nucl-ex]}
  \BibitemShut {NoStop}%
%%CITATION = ARXIV:1505.06924;%%
\bibitem [{\citenamefont {Sick}(2014)}]{Sick:2014yha}%
  \BibitemOpen
  \bibfield  {author} {\bibinfo {author} {\bibfnamefont {I.}~\bibnamefont
  {Sick}},\ }\href {\doibase 10.1103/PhysRevC.90.064002} {\bibfield  {journal}
  {\bibinfo  {journal} {Phys. Rev.}\ }\textbf {\bibinfo {volume} {C90}},\
  \bibinfo {pages} {064002} (\bibinfo {year} {2014})},\ \Eprint
  {http://arxiv.org/abs/1412.2603} {arXiv:1412.2603 [nucl-ex]} \BibitemShut
  {NoStop}%
%%CITATION = ARXIV:1412.2603;%%
\bibitem [{\citenamefont {Gasparian}(2017)}]{Gasparian:2017cgp}%
  \BibitemOpen
  \bibfield  {author} {\bibinfo {author} {\bibfnamefont {A.~H.}\ \bibnamefont
  {Gasparian}} (\bibinfo {collaboration} {PRad}),\ }\bibfield  {booktitle}
  {\emph {\bibinfo {booktitle} {{Proceedings, 14th International Conference on
  Meson-Nucleon Physics and the Structure of the Nucleon (MENU 2016): Kyoto,
  Japan, July 25-30, 2016}}},\ }\href {\doibase 10.7566/JPSCP.13.020052}
  {\bibfield  {journal} {\bibinfo  {journal} {JPS Conf. Proc.}\ }\textbf
  {\bibinfo {volume} {13}},\ \bibinfo {pages} {020052} (\bibinfo {year}
  {2017})}\BibitemShut {NoStop}%
%%CITATION = INSPIRE-1516484;%%
\bibitem [{\citenamefont {Gasparian}(2014)}]{Gasparian:2014rna}%
  \BibitemOpen
  \bibfield  {author} {\bibinfo {author} {\bibfnamefont {A.}~\bibnamefont
  {Gasparian}} (\bibinfo {collaboration} {PRad at JLab}),\ }\bibfield
  {booktitle} {\emph {\bibinfo {booktitle} {{Proceedings, 13th International
  Conference on Meson-Nucleon Physics and the Structure of the Nucleon (MENU
  2013): Rome, Italy, September 30-October 4, 2013}}},\ }\href {\doibase
  10.1051/epjconf/20147307006} {\bibfield  {journal} {\bibinfo  {journal} {EPJ
  Web Conf.}\ }\textbf {\bibinfo {volume} {73}},\ \bibinfo {pages} {07006}
  (\bibinfo {year} {2014})}\BibitemShut {NoStop}%
%%CITATION = 00776,73,07006;%%
\bibitem [{\citenamefont {Meziane}(2013)}]{Meziane:2013yma}%
  \BibitemOpen
  \bibfield  {author} {\bibinfo {author} {\bibfnamefont {M.}~\bibnamefont
  {Meziane}} (\bibinfo {collaboration} {PRad}),\ }\bibfield  {booktitle} {\emph
  {\bibinfo {booktitle} {{Proceedings, Workshop to Explore Physics
  Opportunities with Intense, Polarized Electron Beams up to 300 MeV:
  Cambridge, MA, USA, March 14-16, 2013}}},\ }\href {\doibase
  10.1063/1.4829405} {\bibfield  {journal} {\bibinfo  {journal} {AIP Conf.
  Proc.}\ }\textbf {\bibinfo {volume} {1563}},\ \bibinfo {pages} {183}
  (\bibinfo {year} {2013})}\BibitemShut {NoStop}%
%%CITATION = APCPC,1563,183;%%
\bibitem [{\citenamefont {Fleurbaey}\ \emph {et~al.}(2018)\citenamefont
  {Fleurbaey}, \citenamefont {Galtier}, \citenamefont {Thomas}, \citenamefont
  {Bonnaud}, \citenamefont {Julien}, \citenamefont {Biraben}, \citenamefont
  {Nez}, \citenamefont {Abgrall},\ and\ \citenamefont {Guéna}}]{newH}%
  \BibitemOpen
  \bibfield  {author} {\bibinfo {author} {\bibfnamefont {H.}~\bibnamefont
  {Fleurbaey}}, \bibinfo {author} {\bibfnamefont {S.}~\bibnamefont {Galtier}},
  \bibinfo {author} {\bibfnamefont {S.}~\bibnamefont {Thomas}}, \bibinfo
  {author} {\bibfnamefont {M.}~\bibnamefont {Bonnaud}}, \bibinfo {author}
  {\bibfnamefont {L.}~\bibnamefont {Julien}}, \bibinfo {author} {\bibfnamefont
  {F.}~\bibnamefont {Biraben}}, \bibinfo {author} {\bibfnamefont
  {F.}~\bibnamefont {Nez}}, \bibinfo {author} {\bibfnamefont {M.}~\bibnamefont
  {Abgrall}}, \ and\ \bibinfo {author} {\bibfnamefont {J.}~\bibnamefont
  {Guéna}},\ }\href@noop {} {\  (\bibinfo {year} {2018})},\ \Eprint
  {http://arxiv.org/abs/1801.08816} {1801.08816} \BibitemShut {NoStop}%
\bibitem [{\citenamefont {Liu}\ \emph {et~al.}(2016)\citenamefont {Liu},
  \citenamefont {McKeen},\ and\ \citenamefont {Miller}}]{Liu:2016qwd}%
  \BibitemOpen
  \bibfield  {author} {\bibinfo {author} {\bibfnamefont {Y.-S.}\ \bibnamefont
  {Liu}}, \bibinfo {author} {\bibfnamefont {D.}~\bibnamefont {McKeen}}, \ and\
  \bibinfo {author} {\bibfnamefont {G.~A.}\ \bibnamefont {Miller}},\ }\href
  {\doibase 10.1103/PhysRevLett.117.101801} {\bibfield  {journal} {\bibinfo
  {journal} {Phys. Rev. Lett.}\ }\textbf {\bibinfo {volume} {117}},\ \bibinfo
  {pages} {101801} (\bibinfo {year} {2016})},\ \Eprint
  {http://arxiv.org/abs/1605.04612} {arXiv:1605.04612 [hep-ph]} \BibitemShut
  {NoStop}%
%%CITATION = ARXIV:1605.04612;%%
\bibitem [{\citenamefont {Friar}(1970)}]{Friar:1969zz}%
  \BibitemOpen
  \bibfield  {author} {\bibinfo {author} {\bibfnamefont {J.~L.}\ \bibnamefont
  {Friar}},\ }\href {\doibase 10.1016/0375-9474(70)91112-7} {\bibfield
  {journal} {\bibinfo  {journal} {Nucl. Phys.}\ }\textbf {\bibinfo {volume}
  {A156}},\ \bibinfo {pages} {43} (\bibinfo {year} {1970})}\BibitemShut
  {NoStop}%
%%CITATION = NUPHA,A156,43;%%
\bibitem [{\citenamefont {Friar}\ and\ \citenamefont
  {Gibson}(1978)}]{Friar:1978mr}%
  \BibitemOpen
  \bibfield  {author} {\bibinfo {author} {\bibfnamefont {J.~L.}\ \bibnamefont
  {Friar}}\ and\ \bibinfo {author} {\bibfnamefont {B.~F.}\ \bibnamefont
  {Gibson}},\ }\href {\doibase 10.1103/PhysRevC.18.908} {\bibfield  {journal}
  {\bibinfo  {journal} {Phys. Rev.}\ }\textbf {\bibinfo {volume} {C18}},\
  \bibinfo {pages} {908} (\bibinfo {year} {1978})}\BibitemShut {NoStop}%
%%CITATION = PHRVA,C18,908;%%
\bibitem [{\citenamefont {Coon}\ and\ \citenamefont
  {Barrett}(1987)}]{Coon:1987kt}%
  \BibitemOpen
  \bibfield  {author} {\bibinfo {author} {\bibfnamefont {S.~A.}\ \bibnamefont
  {Coon}}\ and\ \bibinfo {author} {\bibfnamefont {R.~C.}\ \bibnamefont
  {Barrett}},\ }\href {\doibase 10.1103/PhysRevC.36.2189} {\bibfield  {journal}
  {\bibinfo  {journal} {Phys. Rev.}\ }\textbf {\bibinfo {volume} {C36}},\
  \bibinfo {pages} {2189} (\bibinfo {year} {1987})}\BibitemShut {NoStop}%
%%CITATION = PHRVA,C36,2189;%%
\bibitem [{\citenamefont {Miller}\ \emph {et~al.}(1990)\citenamefont {Miller},
  \citenamefont {Nefkens},\ and\ \citenamefont {Slaus}}]{Miller:1990iz}%
  \BibitemOpen
  \bibfield  {author} {\bibinfo {author} {\bibfnamefont {G.~A.}\ \bibnamefont
  {Miller}}, \bibinfo {author} {\bibfnamefont {B.~M.~K.}\ \bibnamefont
  {Nefkens}}, \ and\ \bibinfo {author} {\bibfnamefont {I.}~\bibnamefont
  {Slaus}},\ }\href {\doibase 10.1016/0370-1573(90)90102-8} {\bibfield
  {journal} {\bibinfo  {journal} {Phys. Rept.}\ }\textbf {\bibinfo {volume}
  {194}},\ \bibinfo {pages} {1} (\bibinfo {year} {1990})}\BibitemShut {NoStop}%
%%CITATION = PRPLC,194,1;%%
\bibitem [{\citenamefont {Wiringa}\ \emph {et~al.}(2013)\citenamefont
  {Wiringa}, \citenamefont {Pastore}, \citenamefont {Pieper},\ and\
  \citenamefont {Miller}}]{Wiringa:2013fia}%
  \BibitemOpen
  \bibfield  {author} {\bibinfo {author} {\bibfnamefont {R.~B.}\ \bibnamefont
  {Wiringa}}, \bibinfo {author} {\bibfnamefont {S.}~\bibnamefont {Pastore}},
  \bibinfo {author} {\bibfnamefont {S.~C.}\ \bibnamefont {Pieper}}, \ and\
  \bibinfo {author} {\bibfnamefont {G.~A.}\ \bibnamefont {Miller}},\ }\href
  {\doibase 10.1103/PhysRevC.88.044333} {\bibfield  {journal} {\bibinfo
  {journal} {Phys. Rev.}\ }\textbf {\bibinfo {volume} {C88}},\ \bibinfo {pages}
  {044333} (\bibinfo {year} {2013})},\ \Eprint {http://arxiv.org/abs/1308.5670}
  {arXiv:1308.5670 [nucl-th]} \BibitemShut {NoStop}%
%%CITATION = ARXIV:1308.5670;%%
\bibitem [{\citenamefont {Sick}(2001)}]{Sick:2001rh}%
  \BibitemOpen
  \bibfield  {author} {\bibinfo {author} {\bibfnamefont {I.}~\bibnamefont
  {Sick}},\ }\href {\doibase 10.1016/S0146-6410(01)00156-9} {\bibfield
  {journal} {\bibinfo  {journal} {Prog. Part. Nucl. Phys.}\ }\textbf {\bibinfo
  {volume} {47}},\ \bibinfo {pages} {245} (\bibinfo {year} {2001})},\ \Eprint
  {http://arxiv.org/abs/nucl-ex/0208009} {arXiv:nucl-ex/0208009 [nucl-ex]}
  \BibitemShut {NoStop}%
%%CITATION = NUCL-EX/0208009;%%
\bibitem [{\citenamefont {Juster}\ \emph {et~al.}(1985)\citenamefont {Juster}
  \emph {et~al.}}]{Juster:1985sd}%
  \BibitemOpen
  \bibfield  {author} {\bibinfo {author} {\bibfnamefont {F.~P.}\ \bibnamefont
  {Juster}} \emph {et~al.},\ }\href {\doibase 10.1103/PhysRevLett.55.2261}
  {\bibfield  {journal} {\bibinfo  {journal} {Phys. Rev. Lett.}\ }\textbf
  {\bibinfo {volume} {55}},\ \bibinfo {pages} {2261} (\bibinfo {year}
  {1985})}\BibitemShut {NoStop}%
%%CITATION = PRLTA,55,2261;%%
\bibitem [{\citenamefont {Mccarthy}\ \emph {et~al.}(1977)\citenamefont
  {Mccarthy}, \citenamefont {Sick},\ and\ \citenamefont
  {Whitney}}]{Mccarthy:1977vd}%
  \BibitemOpen
  \bibfield  {author} {\bibinfo {author} {\bibfnamefont {J.~S.}\ \bibnamefont
  {Mccarthy}}, \bibinfo {author} {\bibfnamefont {I.}~\bibnamefont {Sick}}, \
  and\ \bibinfo {author} {\bibfnamefont {R.~R.}\ \bibnamefont {Whitney}},\
  }\href {\doibase 10.1103/PhysRevC.15.1396} {\bibfield  {journal} {\bibinfo
  {journal} {Phys. Rev.}\ }\textbf {\bibinfo {volume} {C15}},\ \bibinfo {pages}
  {1396} (\bibinfo {year} {1977})}\BibitemShut {NoStop}%
%%CITATION = PHRVA,C15,1396;%%
\bibitem [{\citenamefont {Mattuck}(1976)}]{Mattuck:1976xt}%
  \BibitemOpen
  \bibfield  {author} {\bibinfo {author} {\bibfnamefont {R.~D.}\ \bibnamefont
  {Mattuck}},\ }\href@noop {} {\emph {\bibinfo {title} {{A Guide to Feynman
  Diagrams in the Many Body Problem (Second Edition)}}}}\ (\bibinfo {year}
  {1976})\BibitemShut {NoStop}%
%%CITATION = INSPIRE-114145;%%
\bibitem [{\citenamefont {Franke}\ \emph {et~al.}(2017)\citenamefont {Franke},
  \citenamefont {Krauth}, \citenamefont {Antognini}, \citenamefont {Diepold},
  \citenamefont {Kottmann},\ and\ \citenamefont {Pohl}}]{Franke:2017tpc}%
  \BibitemOpen
  \bibfield  {author} {\bibinfo {author} {\bibfnamefont {B.}~\bibnamefont
  {Franke}}, \bibinfo {author} {\bibfnamefont {J.~J.}\ \bibnamefont {Krauth}},
  \bibinfo {author} {\bibfnamefont {A.}~\bibnamefont {Antognini}}, \bibinfo
  {author} {\bibfnamefont {M.}~\bibnamefont {Diepold}}, \bibinfo {author}
  {\bibfnamefont {F.}~\bibnamefont {Kottmann}}, \ and\ \bibinfo {author}
  {\bibfnamefont {R.}~\bibnamefont {Pohl}},\ }\href@noop {} {\  (\bibinfo
  {year} {2017})},\ \Eprint {http://arxiv.org/abs/1705.00352} {arXiv:1705.00352
  [physics.atom-ph]} \BibitemShut {NoStop}%
%%CITATION = ARXIV:1705.00352;%%
\bibitem [{\citenamefont {Leeb}\ and\ \citenamefont
  {Schmiedmayer}(1992)}]{Leeb:1992qf}%
  \BibitemOpen
  \bibfield  {author} {\bibinfo {author} {\bibfnamefont {H.}~\bibnamefont
  {Leeb}}\ and\ \bibinfo {author} {\bibfnamefont {J.}~\bibnamefont
  {Schmiedmayer}},\ }\href {\doibase 10.1103/PhysRevLett.68.1472} {\bibfield
  {journal} {\bibinfo  {journal} {Phys. Rev. Lett.}\ }\textbf {\bibinfo
  {volume} {68}},\ \bibinfo {pages} {1472} (\bibinfo {year}
  {1992})}\BibitemShut {NoStop}%
%%CITATION = PRLTA,68,1472;%%
\bibitem [{\citenamefont {Diepold}\ \emph {et~al.}(2016)\citenamefont
  {Diepold}, \citenamefont {Krauth}, \citenamefont {Franke}, \citenamefont
  {Antognini}, \citenamefont {Kottmann},\ and\ \citenamefont
  {Pohl}}]{Diepold:2016cxv}%
  \BibitemOpen
  \bibfield  {author} {\bibinfo {author} {\bibfnamefont {M.}~\bibnamefont
  {Diepold}}, \bibinfo {author} {\bibfnamefont {J.~J.}\ \bibnamefont {Krauth}},
  \bibinfo {author} {\bibfnamefont {B.}~\bibnamefont {Franke}}, \bibinfo
  {author} {\bibfnamefont {A.}~\bibnamefont {Antognini}}, \bibinfo {author}
  {\bibfnamefont {F.}~\bibnamefont {Kottmann}}, \ and\ \bibinfo {author}
  {\bibfnamefont {R.}~\bibnamefont {Pohl}},\ }\href@noop {} {\  (\bibinfo
  {year} {2016})},\ \Eprint {http://arxiv.org/abs/1606.05231} {arXiv:1606.05231
  [physics.atom-ph]} \BibitemShut {NoStop}%
%%CITATION = ARXIV:1606.05231;%%
\bibitem [{\citenamefont {Pohl}\ \emph {et~al.}(2016)\citenamefont {Pohl} \emph
  {et~al.}}]{Pohl1:2016xoo}%
  \BibitemOpen
  \bibfield  {author} {\bibinfo {author} {\bibfnamefont {R.}~\bibnamefont
  {Pohl}} \emph {et~al.} (\bibinfo {collaboration} {CREMA}),\ }\href {\doibase
  10.1126/science.aaf2468} {\bibfield  {journal} {\bibinfo  {journal}
  {Science}\ }\textbf {\bibinfo {volume} {353}},\ \bibinfo {pages} {669}
  (\bibinfo {year} {2016})}\BibitemShut {NoStop}%
%%CITATION = SCIEA,353,669;%%
\bibitem [{\citenamefont {Krauth}\ \emph {et~al.}(2016)\citenamefont {Krauth},
  \citenamefont {Diepold}, \citenamefont {Franke}, \citenamefont {Antognini},
  \citenamefont {Kottmann},\ and\ \citenamefont {Pohl}}]{Krauth:2015nja}%
  \BibitemOpen
  \bibfield  {author} {\bibinfo {author} {\bibfnamefont {J.~J.}\ \bibnamefont
  {Krauth}}, \bibinfo {author} {\bibfnamefont {M.}~\bibnamefont {Diepold}},
  \bibinfo {author} {\bibfnamefont {B.}~\bibnamefont {Franke}}, \bibinfo
  {author} {\bibfnamefont {A.}~\bibnamefont {Antognini}}, \bibinfo {author}
  {\bibfnamefont {F.}~\bibnamefont {Kottmann}}, \ and\ \bibinfo {author}
  {\bibfnamefont {R.}~\bibnamefont {Pohl}},\ }\href {\doibase
  10.1016/j.aop.2015.12.006} {\bibfield  {journal} {\bibinfo  {journal} {Annals
  Phys.}\ }\textbf {\bibinfo {volume} {366}},\ \bibinfo {pages} {168} (\bibinfo
  {year} {2016})},\ \Eprint {http://arxiv.org/abs/1506.01298} {arXiv:1506.01298
  [physics.atom-ph]} \BibitemShut {NoStop}%
%%CITATION = ARXIV:1506.01298;%%
\bibitem [{\citenamefont {Antognini}\ \emph {et~al.}(2016)\citenamefont
  {Antognini} \emph {et~al.}}]{Antognini:2015moa}%
  \BibitemOpen
  \bibfield  {author} {\bibinfo {author} {\bibfnamefont {A.}~\bibnamefont
  {Antognini}} \emph {et~al.},\ }\bibfield  {booktitle} {\emph {\bibinfo
  {booktitle} {{Proceedings, 21st International Conference on Few-Body Problems
  in Physics (FB21)}}},\ }\href {\doibase 10.1051/epjconf/201611301006}
  {\bibfield  {journal} {\bibinfo  {journal} {EPJ Web Conf.}\ }\textbf
  {\bibinfo {volume} {113}},\ \bibinfo {pages} {01006} (\bibinfo {year}
  {2016})},\ \Eprint {http://arxiv.org/abs/1509.03235} {arXiv:1509.03235
  [physics.atom-ph]} \BibitemShut {NoStop}%
%%CITATION = ARXIV:1509.03235;%%
\bibitem [{\citenamefont {Davoudiasl}\ \emph {et~al.}(2012)\citenamefont
  {Davoudiasl}, \citenamefont {Lee},\ and\ \citenamefont
  {Marciano}}]{Davoudiasl:2012ag}%
  \BibitemOpen
  \bibfield  {author} {\bibinfo {author} {\bibfnamefont {H.}~\bibnamefont
  {Davoudiasl}}, \bibinfo {author} {\bibfnamefont {H.-S.}\ \bibnamefont {Lee}},
  \ and\ \bibinfo {author} {\bibfnamefont {W.~J.}\ \bibnamefont {Marciano}},\
  }\href {\doibase 10.1103/PhysRevD.85.115019} {\bibfield  {journal} {\bibinfo
  {journal} {Phys. Rev.}\ }\textbf {\bibinfo {volume} {D85}},\ \bibinfo {pages}
  {115019} (\bibinfo {year} {2012})},\ \Eprint {http://arxiv.org/abs/1203.2947}
  {arXiv:1203.2947 [hep-ph]} \BibitemShut {NoStop}%
%%CITATION = ARXIV:1203.2947;%%
\bibitem [{\citenamefont {Cloët}\ \emph {et~al.}(2014)\citenamefont {Cloët},
  \citenamefont {Bentz},\ and\ \citenamefont {Thomas}}]{Cloet:2014rja}%
  \BibitemOpen
  \bibfield  {author} {\bibinfo {author} {\bibfnamefont {I.~C.}\ \bibnamefont
  {Cloët}}, \bibinfo {author} {\bibfnamefont {W.}~\bibnamefont {Bentz}}, \
  and\ \bibinfo {author} {\bibfnamefont {A.~W.}\ \bibnamefont {Thomas}},\
  }\href {\doibase 10.1103/PhysRevC.90.045202} {\bibfield  {journal} {\bibinfo
  {journal} {Phys. Rev.}\ }\textbf {\bibinfo {volume} {C90}},\ \bibinfo {pages}
  {045202} (\bibinfo {year} {2014})},\ \Eprint {http://arxiv.org/abs/1405.5542}
  {arXiv:1405.5542 [nucl-th]} \BibitemShut {NoStop}%
%%CITATION = ARXIV:1405.5542;%%
\bibitem [{\citenamefont {Schwinger}(1951)}]{Schwinger:1951nm}%
  \BibitemOpen
  \bibfield  {author} {\bibinfo {author} {\bibfnamefont {J.~S.}\ \bibnamefont
  {Schwinger}},\ }\href {\doibase 10.1103/PhysRev.82.664} {\bibfield  {journal}
  {\bibinfo  {journal} {Phys. Rev.}\ }\textbf {\bibinfo {volume} {82}},\
  \bibinfo {pages} {664} (\bibinfo {year} {1951})}\BibitemShut {NoStop}%
%%CITATION = PHRVA,82,664;%%
\bibitem [{\citenamefont {Ebert}\ \emph {et~al.}(1996)\citenamefont {Ebert},
  \citenamefont {Feldmann},\ and\ \citenamefont {Reinhardt}}]{Ebert:1996vx}%
  \BibitemOpen
  \bibfield  {author} {\bibinfo {author} {\bibfnamefont {D.}~\bibnamefont
  {Ebert}}, \bibinfo {author} {\bibfnamefont {T.}~\bibnamefont {Feldmann}}, \
  and\ \bibinfo {author} {\bibfnamefont {H.}~\bibnamefont {Reinhardt}},\ }\href
  {\doibase 10.1016/0370-2693(96)01158-6} {\bibfield  {journal} {\bibinfo
  {journal} {Phys. Lett. B}\ }\textbf {\bibinfo {volume} {388}},\ \bibinfo
  {pages} {154} (\bibinfo {year} {1996})},\ \Eprint
  {http://arxiv.org/abs/9608223} {arXiv:9608223 [hep-ph]} \BibitemShut
  {NoStop}%
%%CITATION = HEP-PH/9608223;%%
\bibitem [{\citenamefont {Hellstern}\ \emph {et~al.}(1997)\citenamefont
  {Hellstern}, \citenamefont {Alkofer},\ and\ \citenamefont
  {Reinhardt}}]{Hellstern:1997nv}%
  \BibitemOpen
  \bibfield  {author} {\bibinfo {author} {\bibfnamefont {G.}~\bibnamefont
  {Hellstern}}, \bibinfo {author} {\bibfnamefont {R.}~\bibnamefont {Alkofer}},
  \ and\ \bibinfo {author} {\bibfnamefont {H.}~\bibnamefont {Reinhardt}},\
  }\href {\doibase 10.1016/S0375-9474(97)00412-0} {\bibfield  {journal}
  {\bibinfo  {journal} {Nucl. Phys. A}\ }\textbf {\bibinfo {volume} {625}},\
  \bibinfo {pages} {697} (\bibinfo {year} {1997})},\ \Eprint
  {http://arxiv.org/abs/9706551} {arXiv:9706551 [hep-ph]} \BibitemShut
  {NoStop}%
%%CITATION = HEP-PH/9706551;%%
\bibitem [{\citenamefont {Ninomiya}\ \emph {et~al.}(2015)\citenamefont
  {Ninomiya}, \citenamefont {Bentz},\ and\ \citenamefont
  {Cloët}}]{Ninomiya:2014kja}%
  \BibitemOpen
  \bibfield  {author} {\bibinfo {author} {\bibfnamefont {Y.}~\bibnamefont
  {Ninomiya}}, \bibinfo {author} {\bibfnamefont {W.}~\bibnamefont {Bentz}}, \
  and\ \bibinfo {author} {\bibfnamefont {I.~C.}\ \bibnamefont {Cloët}},\
  }\href {\doibase 10.1103/PhysRevC.91.025202} {\bibfield  {journal} {\bibinfo
  {journal} {Phys. Rev.}\ }\textbf {\bibinfo {volume} {C91}},\ \bibinfo {pages}
  {025202} (\bibinfo {year} {2015})},\ \Eprint {http://arxiv.org/abs/1406.7212}
  {arXiv:1406.7212 [nucl-th]} \BibitemShut {NoStop}%
%%CITATION = ARXIV:1406.7212;%%
\bibitem [{\citenamefont {Cao}(2012)}]{Cao:2012nj}%
  \BibitemOpen
  \bibfield  {author} {\bibinfo {author} {\bibfnamefont {F.-G.}\ \bibnamefont
  {Cao}},\ }\href {\doibase 10.1103/PhysRevD.85.057501} {\bibfield  {journal}
  {\bibinfo  {journal} {Phys. Rev.}\ }\textbf {\bibinfo {volume} {D85}},\
  \bibinfo {pages} {057501} (\bibinfo {year} {2012})},\ \Eprint
  {http://arxiv.org/abs/1202.6075} {arXiv:1202.6075 [hep-ph]} \BibitemShut
  {NoStop}%
%%CITATION = ARXIV:1202.6075;%%
\bibitem [{\citenamefont {Bramon}\ \emph {et~al.}(1999)\citenamefont {Bramon},
  \citenamefont {Escribano},\ and\ \citenamefont {Scadron}}]{Bramon:1997va}%
  \BibitemOpen
  \bibfield  {author} {\bibinfo {author} {\bibfnamefont {A.}~\bibnamefont
  {Bramon}}, \bibinfo {author} {\bibfnamefont {R.}~\bibnamefont {Escribano}}, \
  and\ \bibinfo {author} {\bibfnamefont {M.~D.}\ \bibnamefont {Scadron}},\
  }\href {\doibase 10.1007/s100529801009} {\bibfield  {journal} {\bibinfo
  {journal} {Eur. Phys. J.}\ }\textbf {\bibinfo {volume} {C7}},\ \bibinfo
  {pages} {271} (\bibinfo {year} {1999})},\ \Eprint
  {http://arxiv.org/abs/hep-ph/9711229} {arXiv:hep-ph/9711229 [hep-ph]}
  \BibitemShut {NoStop}%
%%CITATION = HEP-PH/9711229;%%
\bibitem [{\citenamefont {Pham}(2015)}]{Pham:2015ina}%
  \BibitemOpen
  \bibfield  {author} {\bibinfo {author} {\bibfnamefont {T.~N.}\ \bibnamefont
  {Pham}},\ }\href {\doibase 10.1103/PhysRevD.92.054021} {\bibfield  {journal}
  {\bibinfo  {journal} {Phys. Rev.}\ }\textbf {\bibinfo {volume} {D92}},\
  \bibinfo {pages} {054021} (\bibinfo {year} {2015})},\ \Eprint
  {http://arxiv.org/abs/1504.05414} {arXiv:1504.05414 [hep-ph]} \BibitemShut
  {NoStop}%
%%CITATION = ARXIV:1504.05414;%%
\bibitem [{\citenamefont {Christ}\ \emph {et~al.}(2010)\citenamefont {Christ},
  \citenamefont {Dawson}, \citenamefont {Izubuchi}, \citenamefont {Jung},
  \citenamefont {Liu}, \citenamefont {Mawhinney}, \citenamefont {Sachrajda},
  \citenamefont {Soni},\ and\ \citenamefont {Zhou}}]{Christ:2010dd}%
  \BibitemOpen
  \bibfield  {author} {\bibinfo {author} {\bibfnamefont {N.~H.}\ \bibnamefont
  {Christ}}, \bibinfo {author} {\bibfnamefont {C.}~\bibnamefont {Dawson}},
  \bibinfo {author} {\bibfnamefont {T.}~\bibnamefont {Izubuchi}}, \bibinfo
  {author} {\bibfnamefont {C.}~\bibnamefont {Jung}}, \bibinfo {author}
  {\bibfnamefont {Q.}~\bibnamefont {Liu}}, \bibinfo {author} {\bibfnamefont
  {R.~D.}\ \bibnamefont {Mawhinney}}, \bibinfo {author} {\bibfnamefont {C.~T.}\
  \bibnamefont {Sachrajda}}, \bibinfo {author} {\bibfnamefont {A.}~\bibnamefont
  {Soni}}, \ and\ \bibinfo {author} {\bibfnamefont {R.}~\bibnamefont {Zhou}},\
  }\href {\doibase 10.1103/PhysRevLett.105.241601} {\bibfield  {journal}
  {\bibinfo  {journal} {Phys. Rev. Lett.}\ }\textbf {\bibinfo {volume} {105}},\
  \bibinfo {pages} {241601} (\bibinfo {year} {2010})},\ \Eprint
  {http://arxiv.org/abs/1002.2999} {arXiv:1002.2999 [hep-lat]} \BibitemShut
  {NoStop}%
%%CITATION = ARXIV:1002.2999;%%
\bibitem [{\citenamefont {Ottnad}\ \emph {et~al.}(2012)\citenamefont {Ottnad},
  \citenamefont {Michael}, \citenamefont {Reker}, \citenamefont {Urbach},
  \citenamefont {Michael}, \citenamefont {Reker},\ and\ \citenamefont
  {Urbach}}]{Ottnad:2012fv}%
  \BibitemOpen
  \bibfield  {author} {\bibinfo {author} {\bibfnamefont {K.}~\bibnamefont
  {Ottnad}}, \bibinfo {author} {\bibfnamefont {C.}~\bibnamefont {Michael}},
  \bibinfo {author} {\bibfnamefont {S.}~\bibnamefont {Reker}}, \bibinfo
  {author} {\bibfnamefont {C.}~\bibnamefont {Urbach}}, \bibinfo {author}
  {\bibfnamefont {C.}~\bibnamefont {Michael}}, \bibinfo {author} {\bibfnamefont
  {S.}~\bibnamefont {Reker}}, \ and\ \bibinfo {author} {\bibfnamefont
  {C.}~\bibnamefont {Urbach}} (\bibinfo {collaboration} {ETM}),\ }\href
  {\doibase 10.1007/JHEP11(2012)048} {\bibfield  {journal} {\bibinfo  {journal}
  {JHEP}\ }\textbf {\bibinfo {volume} {11}},\ \bibinfo {pages} {048} (\bibinfo
  {year} {2012})},\ \Eprint {http://arxiv.org/abs/1206.6719} {arXiv:1206.6719
  [hep-lat]} \BibitemShut {NoStop}%
%%CITATION = ARXIV:1206.6719;%%
\bibitem [{\citenamefont {Gunion}\ \emph {et~al.}(2000)\citenamefont {Gunion},
  \citenamefont {Haber}, \citenamefont {Kane},\ and\ \citenamefont
  {Dawson}}]{Gunion:1989we}%
  \BibitemOpen
  \bibfield  {author} {\bibinfo {author} {\bibfnamefont {J.~F.}\ \bibnamefont
  {Gunion}}, \bibinfo {author} {\bibfnamefont {H.~E.}\ \bibnamefont {Haber}},
  \bibinfo {author} {\bibfnamefont {G.~L.}\ \bibnamefont {Kane}}, \ and\
  \bibinfo {author} {\bibfnamefont {S.}~\bibnamefont {Dawson}},\ }\href@noop {}
  {\bibfield  {journal} {\bibinfo  {journal} {Front. Phys.}\ }\textbf {\bibinfo
  {volume} {80}},\ \bibinfo {pages} {1} (\bibinfo {year} {2000})}\BibitemShut
  {NoStop}%
%%CITATION = FRPHA,80,1;%%
\bibitem [{\citenamefont {Patrignani}\ \emph {et~al.}(2016)\citenamefont
  {Patrignani} \emph {et~al.}}]{Patrignani:2016xqp}%
  \BibitemOpen
  \bibfield  {author} {\bibinfo {author} {\bibfnamefont {C.}~\bibnamefont
  {Patrignani}} \emph {et~al.} (\bibinfo {collaboration} {Particle Data
  Group}),\ }\href {\doibase 10.1088/1674-1137/40/10/100001} {\bibfield
  {journal} {\bibinfo  {journal} {Chin. Phys.}\ }\textbf {\bibinfo {volume}
  {C40}},\ \bibinfo {pages} {100001} (\bibinfo {year} {2016})}\BibitemShut
  {NoStop}%
%%CITATION = CHPHD,C40,100001;%%
\bibitem [{\citenamefont {Asatrian}\ \emph {et~al.}(2012)\citenamefont
  {Asatrian}, \citenamefont {Hovhannisyan},\ and\ \citenamefont
  {Yeghiazaryan}}]{Asatrian:2012tp}%
  \BibitemOpen
  \bibfield  {author} {\bibinfo {author} {\bibfnamefont {H.~M.}\ \bibnamefont
  {Asatrian}}, \bibinfo {author} {\bibfnamefont {A.}~\bibnamefont
  {Hovhannisyan}}, \ and\ \bibinfo {author} {\bibfnamefont {A.}~\bibnamefont
  {Yeghiazaryan}},\ }\href {\doibase 10.1103/PhysRevD.86.114023} {\bibfield
  {journal} {\bibinfo  {journal} {Phys. Rev.}\ }\textbf {\bibinfo {volume}
  {D86}},\ \bibinfo {pages} {114023} (\bibinfo {year} {2012})},\ \Eprint
  {http://arxiv.org/abs/1210.7939} {arXiv:1210.7939 [hep-ph]} \BibitemShut
  {NoStop}%
%%CITATION = ARXIV:1210.7939;%%
\bibitem [{\citenamefont {Bjorken}\ \emph {et~al.}(1988)\citenamefont
  {Bjorken}, \citenamefont {Ecklund}, \citenamefont {Nelson}, \citenamefont
  {Abashian}, \citenamefont {Church}, \citenamefont {Lu}, \citenamefont {Mo},
  \citenamefont {Nunamaker},\ and\ \citenamefont {Rassmann}}]{Bjorken:1988as}%
  \BibitemOpen
  \bibfield  {author} {\bibinfo {author} {\bibfnamefont {J.~D.}\ \bibnamefont
  {Bjorken}}, \bibinfo {author} {\bibfnamefont {S.}~\bibnamefont {Ecklund}},
  \bibinfo {author} {\bibfnamefont {W.~R.}\ \bibnamefont {Nelson}}, \bibinfo
  {author} {\bibfnamefont {A.}~\bibnamefont {Abashian}}, \bibinfo {author}
  {\bibfnamefont {C.}~\bibnamefont {Church}}, \bibinfo {author} {\bibfnamefont
  {B.}~\bibnamefont {Lu}}, \bibinfo {author} {\bibfnamefont {L.~W.}\
  \bibnamefont {Mo}}, \bibinfo {author} {\bibfnamefont {T.~A.}\ \bibnamefont
  {Nunamaker}}, \ and\ \bibinfo {author} {\bibfnamefont {P.}~\bibnamefont
  {Rassmann}},\ }\href {\doibase 10.1103/PhysRevD.38.3375} {\bibfield
  {journal} {\bibinfo  {journal} {Phys. Rev.}\ }\textbf {\bibinfo {volume}
  {D38}},\ \bibinfo {pages} {3375} (\bibinfo {year} {1988})}\BibitemShut
  {NoStop}%
%%CITATION = PHRVA,D38,3375;%%
\bibitem [{\citenamefont {Riordan}\ \emph {et~al.}(1987)\citenamefont {Riordan}
  \emph {et~al.}}]{Riordan:1987aw}%
  \BibitemOpen
  \bibfield  {author} {\bibinfo {author} {\bibfnamefont {E.~M.}\ \bibnamefont
  {Riordan}} \emph {et~al.},\ }\href {\doibase 10.1103/PhysRevLett.59.755}
  {\bibfield  {journal} {\bibinfo  {journal} {Phys. Rev. Lett.}\ }\textbf
  {\bibinfo {volume} {59}},\ \bibinfo {pages} {755} (\bibinfo {year}
  {1987})}\BibitemShut {NoStop}%
%%CITATION = PRLTA,59,755;%%
\bibitem [{\citenamefont {Davier}\ and\ \citenamefont
  {Nguyen~Ngoc}(1989)}]{Davier:1989wz}%
  \BibitemOpen
  \bibfield  {author} {\bibinfo {author} {\bibfnamefont {M.}~\bibnamefont
  {Davier}}\ and\ \bibinfo {author} {\bibfnamefont {H.}~\bibnamefont
  {Nguyen~Ngoc}},\ }\href {\doibase 10.1016/0370-2693(89)90174-3} {\bibfield
  {journal} {\bibinfo  {journal} {Phys. Lett.}\ }\textbf {\bibinfo {volume}
  {B229}},\ \bibinfo {pages} {150} (\bibinfo {year} {1989})}\BibitemShut
  {NoStop}%
%%CITATION = PHLTA,B229,150;%%
\bibitem [{\citenamefont {Pospelov}(2009)}]{Pospelov:2008zw}%
  \BibitemOpen
  \bibfield  {author} {\bibinfo {author} {\bibfnamefont {M.}~\bibnamefont
  {Pospelov}},\ }\href {\doibase 10.1103/PhysRevD.80.095002} {\bibfield
  {journal} {\bibinfo  {journal} {Phys. Rev.}\ }\textbf {\bibinfo {volume}
  {D80}},\ \bibinfo {pages} {095002} (\bibinfo {year} {2009})},\ \Eprint
  {http://arxiv.org/abs/0811.1030} {arXiv:0811.1030 [hep-ph]} \BibitemShut
  {NoStop}%
%%CITATION = ARXIV:0811.1030;%%
\bibitem [{\citenamefont {Bouchendira}\ \emph {et~al.}(2011)\citenamefont
  {Bouchendira}, \citenamefont {Clade}, \citenamefont {Guellati-Khelifa},
  \citenamefont {Nez},\ and\ \citenamefont {Biraben}}]{Bouchendira:2010es}%
  \BibitemOpen
  \bibfield  {author} {\bibinfo {author} {\bibfnamefont {R.}~\bibnamefont
  {Bouchendira}}, \bibinfo {author} {\bibfnamefont {P.}~\bibnamefont {Clade}},
  \bibinfo {author} {\bibfnamefont {S.}~\bibnamefont {Guellati-Khelifa}},
  \bibinfo {author} {\bibfnamefont {F.}~\bibnamefont {Nez}}, \ and\ \bibinfo
  {author} {\bibfnamefont {F.}~\bibnamefont {Biraben}},\ }\href {\doibase
  10.1103/PhysRevLett.106.080801} {\bibfield  {journal} {\bibinfo  {journal}
  {Phys. Rev. Lett.}\ }\textbf {\bibinfo {volume} {106}},\ \bibinfo {pages}
  {080801} (\bibinfo {year} {2011})},\ \Eprint {http://arxiv.org/abs/1012.3627}
  {arXiv:1012.3627 [physics.atom-ph]} \BibitemShut {NoStop}%
%%CITATION = ARXIV:1012.3627;%%
\bibitem [{\citenamefont {Tsertos}\ \emph {et~al.}(1989)\citenamefont
  {Tsertos}, \citenamefont {Kozhuharov}, \citenamefont {Armbruster},
  \citenamefont {Kienle}, \citenamefont {Krusche},\ and\ \citenamefont
  {Schreckenbach}}]{Tsertos:1989gv}%
  \BibitemOpen
  \bibfield  {author} {\bibinfo {author} {\bibfnamefont {H.}~\bibnamefont
  {Tsertos}}, \bibinfo {author} {\bibfnamefont {C.}~\bibnamefont {Kozhuharov}},
  \bibinfo {author} {\bibfnamefont {P.}~\bibnamefont {Armbruster}}, \bibinfo
  {author} {\bibfnamefont {P.}~\bibnamefont {Kienle}}, \bibinfo {author}
  {\bibfnamefont {B.}~\bibnamefont {Krusche}}, \ and\ \bibinfo {author}
  {\bibfnamefont {K.}~\bibnamefont {Schreckenbach}},\ }\href {\doibase
  10.1103/PhysRevD.40.1397} {\bibfield  {journal} {\bibinfo  {journal} {Phys.
  Rev.}\ }\textbf {\bibinfo {volume} {D40}},\ \bibinfo {pages} {1397} (\bibinfo
  {year} {1989})}\BibitemShut {NoStop}%
%%CITATION = PHRVA,D40,1397;%%
\bibitem [{\citenamefont {Miller}\ \emph {et~al.}(2011)\citenamefont {Miller},
  \citenamefont {Thomas}, \citenamefont {Carroll},\ and\ \citenamefont
  {Rafelski}}]{Miller:2011yw}%
  \BibitemOpen
  \bibfield  {author} {\bibinfo {author} {\bibfnamefont {G.~A.}\ \bibnamefont
  {Miller}}, \bibinfo {author} {\bibfnamefont {A.~W.}\ \bibnamefont {Thomas}},
  \bibinfo {author} {\bibfnamefont {J.~D.}\ \bibnamefont {Carroll}}, \ and\
  \bibinfo {author} {\bibfnamefont {J.}~\bibnamefont {Rafelski}},\ }\href
  {\doibase 10.1103/PhysRevA.84.020101} {\bibfield  {journal} {\bibinfo
  {journal} {Phys. Rev.}\ }\textbf {\bibinfo {volume} {A84}},\ \bibinfo {pages}
  {020101} (\bibinfo {year} {2011})},\ \Eprint {http://arxiv.org/abs/1101.4073}
  {arXiv:1101.4073 [physics.atom-ph]} \BibitemShut {NoStop}%
%%CITATION = ARXIV:1101.4073;%%
\bibitem [{\citenamefont {Miller}\ \emph {et~al.}(2012)\citenamefont {Miller},
  \citenamefont {Thomas},\ and\ \citenamefont {Carroll}}]{Miller:2012ht}%
  \BibitemOpen
  \bibfield  {author} {\bibinfo {author} {\bibfnamefont {G.~A.}\ \bibnamefont
  {Miller}}, \bibinfo {author} {\bibfnamefont {A.~W.}\ \bibnamefont {Thomas}},
  \ and\ \bibinfo {author} {\bibfnamefont {J.~D.}\ \bibnamefont {Carroll}},\
  }\href {\doibase 10.1103/PhysRevC.86.065201} {\bibfield  {journal} {\bibinfo
  {journal} {Phys. Rev.}\ }\textbf {\bibinfo {volume} {C86}},\ \bibinfo {pages}
  {065201} (\bibinfo {year} {2012})},\ \Eprint {http://arxiv.org/abs/1207.0549}
  {arXiv:1207.0549 [nucl-th]} \BibitemShut {NoStop}%
%%CITATION = ARXIV:1207.0549;%%
\bibitem [{\citenamefont {Miller}(2013)}]{Miller:2012ne}%
  \BibitemOpen
  \bibfield  {author} {\bibinfo {author} {\bibfnamefont {G.~A.}\ \bibnamefont
  {Miller}},\ }\href {\doibase 10.1016/j.physletb.2012.11.016} {\bibfield
  {journal} {\bibinfo  {journal} {Phys. Lett.}\ }\textbf {\bibinfo {volume}
  {B718}},\ \bibinfo {pages} {1078} (\bibinfo {year} {2013})},\ \Eprint
  {http://arxiv.org/abs/1209.4667} {arXiv:1209.4667 [nucl-th]} \BibitemShut
  {NoStop}%
%%CITATION = ARXIV:1209.4667;%%
\bibitem [{\citenamefont {Eides}\ \emph {et~al.}(2001)\citenamefont {Eides},
  \citenamefont {Grotch},\ and\ \citenamefont {Shelyuto}}]{Eides:2000xc}%
  \BibitemOpen
  \bibfield  {author} {\bibinfo {author} {\bibfnamefont {M.~I.}\ \bibnamefont
  {Eides}}, \bibinfo {author} {\bibfnamefont {H.}~\bibnamefont {Grotch}}, \
  and\ \bibinfo {author} {\bibfnamefont {V.~A.}\ \bibnamefont {Shelyuto}},\
  }\href {\doibase 10.1016/S0370-1573(00)00077-6} {\bibfield  {journal}
  {\bibinfo  {journal} {Phys. Rept.}\ }\textbf {\bibinfo {volume} {342}},\
  \bibinfo {pages} {63} (\bibinfo {year} {2001})},\ \Eprint
  {http://arxiv.org/abs/hep-ph/0002158} {arXiv:hep-ph/0002158 [hep-ph]}
  \BibitemShut {NoStop}%
%%CITATION = HEP-PH/0002158;%%
\bibitem [{\citenamefont {Merkel}\ \emph {et~al.}(2014)\citenamefont {Merkel}
  \emph {et~al.}}]{Merkel:2014avp}%
  \BibitemOpen
  \bibfield  {author} {\bibinfo {author} {\bibfnamefont {H.}~\bibnamefont
  {Merkel}} \emph {et~al.},\ }\href {\doibase 10.1103/PhysRevLett.112.221802}
  {\bibfield  {journal} {\bibinfo  {journal} {Phys. Rev. Lett.}\ }\textbf
  {\bibinfo {volume} {112}},\ \bibinfo {pages} {221802} (\bibinfo {year}
  {2014})},\ \Eprint {http://arxiv.org/abs/1404.5502} {arXiv:1404.5502
  [hep-ex]} \BibitemShut {NoStop}%
%%CITATION = ARXIV:1404.5502;%%
\bibitem [{\citenamefont {Lees}\ \emph {et~al.}(2014)\citenamefont {Lees} \emph
  {et~al.}}]{Lees:2014xha}%
  \BibitemOpen
  \bibfield  {author} {\bibinfo {author} {\bibfnamefont {J.~P.}\ \bibnamefont
  {Lees}} \emph {et~al.} (\bibinfo {collaboration} {BaBar}),\ }\href {\doibase
  10.1103/PhysRevLett.113.201801} {\bibfield  {journal} {\bibinfo  {journal}
  {Phys. Rev. Lett.}\ }\textbf {\bibinfo {volume} {113}},\ \bibinfo {pages}
  {201801} (\bibinfo {year} {2014})},\ \Eprint {http://arxiv.org/abs/1406.2980}
  {arXiv:1406.2980 [hep-ex]} \BibitemShut {NoStop}%
%%CITATION = ARXIV:1406.2980;%%
\bibitem [{\citenamefont {Gan}(2017)}]{Gan:2017kfr}%
  \BibitemOpen
  \bibfield  {author} {\bibinfo {author} {\bibfnamefont {L.}~\bibnamefont
  {Gan}},\ }\bibfield  {booktitle} {\emph {\bibinfo {booktitle} {{Proceedings,
  14th International Conference on Meson-Nucleon Physics and the Structure of
  the Nucleon (MENU 2016): Kyoto, Japan, July 25-30, 2016}}},\ }\href {\doibase
  10.7566/JPSCP.13.020063} {\bibfield  {journal} {\bibinfo  {journal} {JPS
  Conf. Proc.}\ }\textbf {\bibinfo {volume} {13}},\ \bibinfo {pages} {020063}
  (\bibinfo {year} {2017})}\BibitemShut {NoStop}%
%%CITATION = INSPIRE-1516490;%%
\bibitem [{\citenamefont {Gatto}\ \emph {et~al.}(2016)\citenamefont {Gatto},
  \citenamefont {Fabela~Enriquez},\ and\ \citenamefont
  {Pedraza~Morales}}]{Gatto:2016rae}%
  \BibitemOpen
  \bibfield  {author} {\bibinfo {author} {\bibfnamefont {C.}~\bibnamefont
  {Gatto}}, \bibinfo {author} {\bibfnamefont {B.}~\bibnamefont
  {Fabela~Enriquez}}, \ and\ \bibinfo {author} {\bibfnamefont {M.~I.}\
  \bibnamefont {Pedraza~Morales}} (\bibinfo {collaboration} {REDTOP}),\
  }\bibfield  {booktitle} {\emph {\bibinfo {booktitle} {{Proceedings, 38th
  International Conference on High Energy Physics (ICHEP 2016): Chicago, IL,
  USA, August 3-10, 2016}}},\ }\href@noop {} {\bibfield  {journal} {\bibinfo
  {journal} {PoS}\ }\textbf {\bibinfo {volume} {ICHEP2016}},\ \bibinfo {pages}
  {812} (\bibinfo {year} {2016})}\BibitemShut {NoStop}%
%%CITATION = POSCI,ICHEP2016,812;%%
\bibitem [{\citenamefont {Gilman}\ \emph {et~al.}(2013)\citenamefont {Gilman}
  \emph {et~al.}}]{Gilman:2013eiv}%
  \BibitemOpen
  \bibfield  {author} {\bibinfo {author} {\bibfnamefont {R.}~\bibnamefont
  {Gilman}} \emph {et~al.} (\bibinfo {collaboration} {MUSE}),\ }\href@noop {}
  {\  (\bibinfo {year} {2013})},\ \Eprint {http://arxiv.org/abs/1303.2160}
  {arXiv:1303.2160 [nucl-ex]} \BibitemShut {NoStop}%
%%CITATION = ARXIV:1303.2160;%%
\bibitem [{\citenamefont {Chapelain}(2017)}]{Chapelain:2017syu}%
  \BibitemOpen
  \bibfield  {author} {\bibinfo {author} {\bibfnamefont {A.}~\bibnamefont
  {Chapelain}} (\bibinfo {collaboration} {Muon g-2}),\ }\bibfield  {booktitle}
  {\emph {\bibinfo {booktitle} {{Proceedings, 12th Conference on Quark
  Confinement and the Hadron Spectrum (Confinement XII): Thessaloniki,
  Greece}}},\ }\href {\doibase 10.1051/epjconf/201713708001} {\bibfield
  {journal} {\bibinfo  {journal} {EPJ Web Conf.}\ }\textbf {\bibinfo {volume}
  {137}},\ \bibinfo {pages} {08001} (\bibinfo {year} {2017})},\ \Eprint
  {http://arxiv.org/abs/1701.02807} {arXiv:1701.02807 [physics.ins-det]}
  \BibitemShut {NoStop}%
%%CITATION = ARXIV:1701.02807;%%
\bibitem [{\citenamefont {Iinuma}\ \emph {et~al.}(2016)\citenamefont {Iinuma},
  \citenamefont {Nakayama}, \citenamefont {Oide}, \citenamefont {Sasaki},
  \citenamefont {Saito}, \citenamefont {Mibe},\ and\ \citenamefont
  {Abe}}]{Iinuma:2016zfu}%
  \BibitemOpen
  \bibfield  {author} {\bibinfo {author} {\bibfnamefont {H.}~\bibnamefont
  {Iinuma}}, \bibinfo {author} {\bibfnamefont {H.}~\bibnamefont {Nakayama}},
  \bibinfo {author} {\bibfnamefont {K.}~\bibnamefont {Oide}}, \bibinfo {author}
  {\bibfnamefont {K.-i.}\ \bibnamefont {Sasaki}}, \bibinfo {author}
  {\bibfnamefont {N.}~\bibnamefont {Saito}}, \bibinfo {author} {\bibfnamefont
  {T.}~\bibnamefont {Mibe}}, \ and\ \bibinfo {author} {\bibfnamefont
  {M.}~\bibnamefont {Abe}},\ }\href {\doibase 10.1016/j.nima.2016.05.126}
  {\bibfield  {journal} {\bibinfo  {journal} {Nucl. Instrum. Meth.}\ }\textbf
  {\bibinfo {volume} {A832}},\ \bibinfo {pages} {51} (\bibinfo {year}
  {2016})}\BibitemShut {NoStop}%
%%CITATION = NUIMA,A832,51;%%
\bibitem [{Den()}]{Denisov:2018aa}%
  \BibitemOpen
  \bibinfo {note} {{O. Denisov, private communication}}\BibitemShut {NoStop}%
\end{thebibliography}%

\end{document}